\documentclass[pra,twocolumn,superscriptaddress]{revtex4-2}

\usepackage{amsmath}
\usepackage{amssymb}

\newcommand{\lsite}{\mvec}
\newcommand{\ksite}{\nvec}

\newcommand{\rvec}{\boldsymbol{r}}
\newcommand{\ivec}{\boldsymbol{i}}
\newcommand{\rvecd}{{\boldsymbol{r}}^{\scriptscriptstyle\parallel}}

\newcommand{\Rvec}{\boldsymbol{R}}
\newcommand{\Rvecd}{{\boldsymbol{R}}^{\scriptscriptstyle\parallel}}

\newcommand{\avec}{\boldsymbol{a}}
\newcommand{\lvec}{\boldsymbol{l}}
\newcommand{\nvec}{\boldsymbol{n}}
\newcommand{\mvec}{\boldsymbol{m}}
\newcommand{\kvec}{\boldsymbol{k}}

\newcommand{\drm}{\mathrm{d}}

\newcommand{\omegaza}{\omega_{z,\ksite}}

\newcommand{\omegaxya}{\omega_{xy,\ksite}}

\newcommand{\zra}{z_{\ksite}}
\newcommand{\zrb}{z_{\lsite}}
\newcommand{\zr}{z_{R}}
\newcommand{\vzpan}{V_{0,\rm p}}
\newcommand{\vzspot}{V_{0,\rm s}}
\newcommand{\vzspota}{V_{\ksite}}
\newcommand{\vzspotb}{V_{\lsite}}
\newcommand{\wpan}{\textrm{w}_{\rm p}}
\newcommand{\wspot}{\textrm{w}_{\rm 0}}
\newcommand{\wspota}{\textrm{w}_{\ksite}}
\newcommand{\wspotb}{\textrm{w}_{\lsite}}

\usepackage{graphicx}

\begin{document}

%\title{Multi-band Hubbard models and impurity problems in programmable optical lattices}
% Arbitrary Hubbard models in programmable optical lattices
% 
\title{Hubbard models with arbitrary structures in programmable optical lattices}

\author{J.P. Hague}
\affiliation{School of Physical Sciences, The Open University, Walton Hall, Milton Keynes, MK7 6AA, UK}

\author{L. Petit}
\affiliation{Science and Technology Facilities Council, Daresbury Laboratory, Daresbury WA4 4AD, UK}

\author{C. MacCormick}
\affiliation{School of Physical Sciences, The Open University, Walton Hall, Milton Keynes, MK7 6AA, UK}

\begin{abstract}
We investigate the use of programmable optical lattices for quantum simulation of Hubbard models, determining analytic expressions for the hopping and Hubbard $U$, finding that they are suitable for emulating strongly correlated systems with arbitrary structures, including those with multiple site basis and impurities. Programmable potentials are highly flexible, with the ability to control the depth and shape of individual sites in the optical lattice dynamically. Quantum simulators of Hubbard models with (1) arbitrary basis are required to represent many real materials of contemporary interest, (2) broken translational symmetry are needed to study impurity physics, and (3) dynamical lattices are needed to investigate strong correlation out of equilibrium. We derive analytic expressions for Hubbard Hamiltonians in programmable potential systems. We find experimental parameters for quantum simulation of Hubbard models with arbitrary basis, concluding that programmable optical lattices are suitable for this purpose. We discuss how programmable optical lattices can be used for quantum simulation of dynamical multi-band Hubbard models that represent complicated compounds, impurities, and non-equilibrium physics.
\end{abstract}

\maketitle

\section{Introduction}

Programmable potentials are an advanced paradigm for the formation of optical lattices, with applications in quantum technologies such as quantum simulation and quantum computing \cite{henderson2009,ebadi2020}. Key experimental realizations of programmable potentials use acousto-optic modulators (AOMs) \cite{henderson2009}, and holographic techniques \cite{nogrette2014,ebadi2020,Barredo2016,Barredo2018}, to form programmable quantum simulators with bespoke optical lattices. Typical programmable potential systems confine cold atoms to a horizontal plane using dynamic optical tweezers. In this way it is possible for the user to construct arbitrary potentials such as rings from individual Gaussian spots \cite{henderson2009}. Holographic arrays can also be used as the optical tweezers \cite{nogrette2014,Barredo2018,ebadi2020}. Painted potential systems are highly tunable, and the properties of individual lattice sites can be addressed by changing beam waist and spot depth. Programmable potentials can be dynamical, break translational symmetry, and have arbitrary patterns that represent basis. 

Major successes of cold-atom quantum simulators include the emulation of standard models of strong correlation: the single-band Bose--Hubbard \cite{greiner2002} and Fermi--Hubbard models \cite{jordens2008}. Quantum simulators emulate models using highly controllable systems such as cold atoms to provide insight into the behavior of complicated condensed matter systems. The standard approach is to form static sinusoidal optical lattices representing simple crystal structures using counterpropagating beams. Cold atoms are then loaded into the lattice. This results in a single-band Hubbard model controlled using the depth of a sinusoidal potential and scattering length of a Feshbach resonance \cite{bloch2008a}.  The high level of control over the Hubbard parameters has allowed the direct observation of superfluid--insulator and metal--insulator transitions \cite{greiner2002, jordens2008}. However, sinusoidal optical lattices are difficult to generalize for the quantum simulation of Hubbard models with arbitrary basis, or that break translational symmetry.

Hubbard models with a complicated basis are required to represent many real materials of contemporary interest (in the following, we shall use complicated and arbitrary interchangably). Most real low-dimensional materials have a basis containing atoms of different species. For example, a key element of cuprate superconductors is CuO$_{2}$ layers \cite{bednorz1986}. Graphene and other atomically thick van der Walls materials have a basis of two (or more) atoms per unit cell \cite{novoselov2004}. This means that models (and thus quantum simulators) of these materials require a basis of sites that is controllable and extensible. Such a basis can lead to multiple interacting bands.  Moreover, in quantum materials with impurities, translational symmetry is broken. Quantum simulators  of impurities are impossible to construct using purely sinusoidal optical lattices - such lattices must be augmented by some additional optical structure, such as an overlapped but incommensurate lattice \cite{Fallani2007}, or with an additional laser speckle pattern \cite{Roati2008} (these approaches are reviewed in \cite{SanchezPalencia2009}). Impurities are of interest because they can lead to radically different behavior such as the Kondo effect \cite{andrei1983}. There is also interest in non-equilibrium phenomena such as quenches, periodic driving, the dynamics of quantum phase transitions, and transport \cite{eisert2015}.  Implementations of quantum simulators for such systems would be of high interest for both the quantum simulation and condensed matter communities. 

Our aim is to determine how complicated Hubbard models can be emulated using cold atoms in programmable potentials. Programmable potentials have suitable properties for investigation of arbitrary Hubbard models, namely high control over spatial properties of the potential to introduce arbitrary basis, the ability to break translational symmetry, and dynamical potentials to push systems out of equilibrium. The Hubbard model is a standard model of condensed matter \cite{hubbard1963}, written in its generalized extended form as \cite{hirsch1984},
\begin{align}
\hat{H} = & \sum_{\nvec\sigma}E_{\nvec}c^{\dagger}_{\nvec\sigma}c_{\nvec\sigma}-\sum_{\nvec\mvec\sigma}t_{\nvec\mvec} c^{\dagger}_{\nvec\sigma}c_{\mvec\sigma}\nonumber\\
& \hspace{10mm}+ \sum_{\nvec}U_{\nvec\nvec}n_{\nvec\uparrow}n_{\nvec\downarrow} + \sum_{\nvec\mvec}U_{\nvec\mvec}n_{\nvec}n_{\mvec}
\end{align}
where $\nvec$ and $\mvec$ are indices to lattice sites, $c^{\dagger}_{k\sigma}$ ($c_{k\sigma}$) are creation (annihilation) operators for an electron at site $\nvec$ with spin $\sigma$, $n_{\nvec}=n_{\nvec\uparrow}+n_{\nvec\downarrow}$ and $n_{\nvec\sigma}=c^{\dagger}_{\nvec\sigma}c_{\nvec\sigma}$ the corresponding number operator. Each site can have its own energy, $E_{\nvec}$, and Coulomb repulsion, $U_{\nvec\nvec}$. Interaction between sites is denoted $U_{\nvec\mvec}$. We note that the hopping parameter, $t$, is often denoted $J$ in the  quantum simulator literature. We allow for the possibility that sites are not equivalent. In such cases the Hamiltonian can represent impurity problems or has a basis leading to multiple bands.

There are several non-square lattices that can be implemented using standard systems on counter-propagating beams. For example honeycomb lattices \cite{tarruell2012}, triangular lattices \cite{struck2011}, Kagome lattices \cite{jo2012}, double-well \cite{sebbystrabley2006} and periodically driven (Floquet engineered) optical lattices \cite{goldman2016}. Each of these lattices must be set up individually with painstaking experimental effort. These are typically lattices with a single type of site, and do not generally reflect the complicated basis of atoms found in many low dimensional materials (for example many low dimensional van der Waals materials related to graphene have multiple atoms of different types per lattice site). Although more challenging to set up, programmable systems using optical tweezers can be switched from one lattice type to another instantaneously, and there can be very complex arrangements of sites with different types, more closely relating to the situation in real low-dimensional materials \cite{henderson2009}. There are also classic problems in condensed matter physics, such as the Kondo problem \cite{kondo1964}, where translational symmetry is broken leading to qualitatively different behavior, which would be very challenging to emulate using systems of counter-propagating beams.

To our knowledge, no analytic calculations of hopping, $t$, and Hubbard $U$ have been made specifically for programmable optical lattice systems. Such lattices consist of separate and independently controllable finite Gaussian wells (a form that emulates the nuclear potentials in condensed matter systems where each atom can have its own atomic number).  Lattices in programmable quantum simulators tend to have large inter-site spacing, potentially leading to quantum simulation with lower energy scales, so it is of value to establish if this places limits on quantum simulation of strong correlation problems. Existing analytic estimates for the hopping $t$  and Hubbard $U$ of cold-atom quantum simulators typically relate to cold atoms moving in sinusoidal potentials \cite{jaksch1998,bloch2008a}. We note that Wall et al. have carried out numerical calculations for systems of optical tweezers, providing a semi-analytic expression by fitting to the hopping for the specific case of two identical sites \cite{wall2015}.

Our goal in this paper is to derive analytic estimates of the parameters of arbitrary Hubbard models with basis and translational symmetry breaking that are valid for programmable optical lattices and use these to identify how such models can be implemented as quantum simulators. This paper is organized as follows. In Sec. \ref{sec:potential} we state the form of the optical lattice potential. In Sec. \ref{sec:deepwells}, approximations for the Hubbard model resulting from painted potentials are derived for the case where wells are deep. We discuss the relevance to experimental implementations in Sec. \ref{sec:implementation}.  Finally we discuss applications in Sec. \ref{sec:discussion}. To assist with the extensive notation in this article, we summarize the meaning of all symbols within a table in the Appendix.

\section{Programmable potentials}
\label{sec:potential}

Programmable potentials, formed using either acousto-optic modulators \cite{henderson2009} or holographic systems \cite{nogrette2014,Barredo2016,Barredo2018,ebadi2020} allow a high level of control over the form of optical lattices.  Cold atoms irradiated with far-detuned light of intensity $I(\rvec)$ experience a potential,
\begin{equation}
  V_{\rm dip}(\rvec) = \frac{3\pi c^2}{2\omega_0^3}\frac{\Gamma}{\Delta} I(\rvec) = \frac{3\pi\lambda_{0}^3}{2 c}\frac{\Gamma}{\Delta} I(\rvec).
  \label{eqn:vdipfromintensity}
\end{equation}

The detuning parameter, $\Delta =\omega_{\mathrm{Las}} - \omega_0 $, represents the detuning of a laser with frequency $\omega_{\mathrm{Las}}$ from the transition frequency $\omega_0$, and is important for determining if the potential in Eqn. \ref{eqn:vdipfromintensity} is attractive (atoms are drawn to regions of higher laser intensity) or repulsive (atoms avoid regions of high laser intensity). We assume red detuning throughout, where the wavelength of the laser beam $\lambda_{\mathrm{Las}}>\lambda_0$, such that  $\Delta = \omega_{\mathrm{Las}} - \omega_0 <0$ and Eqn.~(\ref{eqn:vdipfromintensity}) represents an attractive potential,
$\lambda_{0}$ is the corresponding transition wavelength, $c$ the speed of light, and $\Gamma$ is lifetime of the transition \cite{grimm2000}.

At the core of a typical programmable optical lattice is a flat optical pancake that confines cold atoms to a quasi-two-dimensional region of space. The optical pancake is formed by focusing a beam to make a disc of thickness $\sim 10\,\mu$m and radius $\sim 0.5\,$mm. Within the region we shall be interested in, the properties of the pancake are constant. The purpose of the optical pancake is to ensure that atoms are confined within a two-dimensional plane so that they can be trapped reliably by deeper potentials provided by the optical tweezers. Without the pancake the atoms sag due to the force of gravity. 
Beyond this practical consideration, the pancake does not strongly affect the properties of the quantum simulator. We note that recently, three-dimensional atomic arrays have been created in the absence of an optical pancake \cite{Barredo2018}, however there was no tunneling between lattice sites and so a supporting optical trap was not required, so a 3D equivalent to the pancake would still be required.

We use a Gaussian approximation for the shape of the optical pancake towards its center,
\begin{equation}
  V_{\rm pan}(z) = -\vzpan \exp(-2z^2/\wpan^2)
\end{equation}
where $\vzpan$ is the magnitude of the pancake potential at $z=0$, $\wpan$ is the waist of the optical pancake. The pancake varies slowly towards its center, but has spatial dependence towards its edge. Later in this paper, we will identify conditions for neglecting the pancake potential.

The optical pancake is punctured by Gaussian beams that are applied roughly perpendicular to the pancake to form lattice sites. We refer to these sites as spots. The spot potential is formed from a Gaussian beam, which has the form,
\begin{equation}
  V_{\rm spot}(\rvecd,z) = -\frac{\vzspot}{[\bar{\textrm{w}}(z)]^2}\exp\left(-\frac{2 |\rvecd|^2}{\wspot^2 [\bar{\textrm{w}}(z)]^2}\right)
  \label{eqn:singlesplotpotenial}
\end{equation}
where $\wspot$ is the waist of the Gaussian beam at $z=0$, and $\bar{\textrm{w}}(z)=\sqrt{1+(z/z_{R})^2}$ (see e.g. \cite{bloch2008a}). The parameter $\vzspot$ can be determined by comparing Eqn. \ref{eqn:vdipfromintensity},  to the intensity of a Gaussian beam propagating along the $z$-direction,
\begin{equation}
I(\rvecd,z)=\frac{2P_{N}}{\pi [\wspot \bar{\textrm{w}}(z)]^2}\exp\left(-\frac{2 |\rvecd|^2}{[\wspot \bar{\textrm{w}}(z)]^2}\right),
\label{eqn:intensity}
\end{equation} 
leading to,
\begin{equation}
\vzspot=\frac{3\pi\lambda_{0}^3}{2 c}\frac{\Gamma}{\Delta} \frac{2P_{N}}{\pi \wspot^2},
\label{eqn:vdipfromtransition}
\end{equation}
where $P_{N} = P/N$ is the power of the beam, $P$, distributed between $N$ lattice sites; $z_{R}=\pi
\wspot^2/\lambda_{\rm Las}$ is the Rayleigh length. Spots are painted towards the center of the pancake, so that on the lengthscale between spots, the pancake potential varies slowly.

The total potential experienced by cold atoms in the quantum simulator has the form,
\begin{equation}
V(\rvecd,z) = V_{\rm pan}(z)+\sum_{\lsite}V_{\rm spot,\lsite}(\rvecd,z)
\label{eqn:totalpotential}
\end{equation}
where we have used the shorthand $V_{\rm spot,\lsite}(\rvecd,z)=V_{\rm spot}(\rvecd-\Rvecd_{\lsite},z)$ for a spot centered about position $\Rvecd_{\lsite}$ with depth $\vzspotb$ and waist $\wspotb$. The potential is highly anisotropic.  Since the system is anisotropic, we denote vectors within the pancake with $\parallel$. So $\rvecd$ is a vector that lies within the plane of the optical pancake, $\Rvecd_{i}$ is a two-dimensional vector to the centers of the Gaussian wells, and the $z$-axis is perpendicular to the pancake. 

\begin{figure}
\includegraphics[width=85mm]{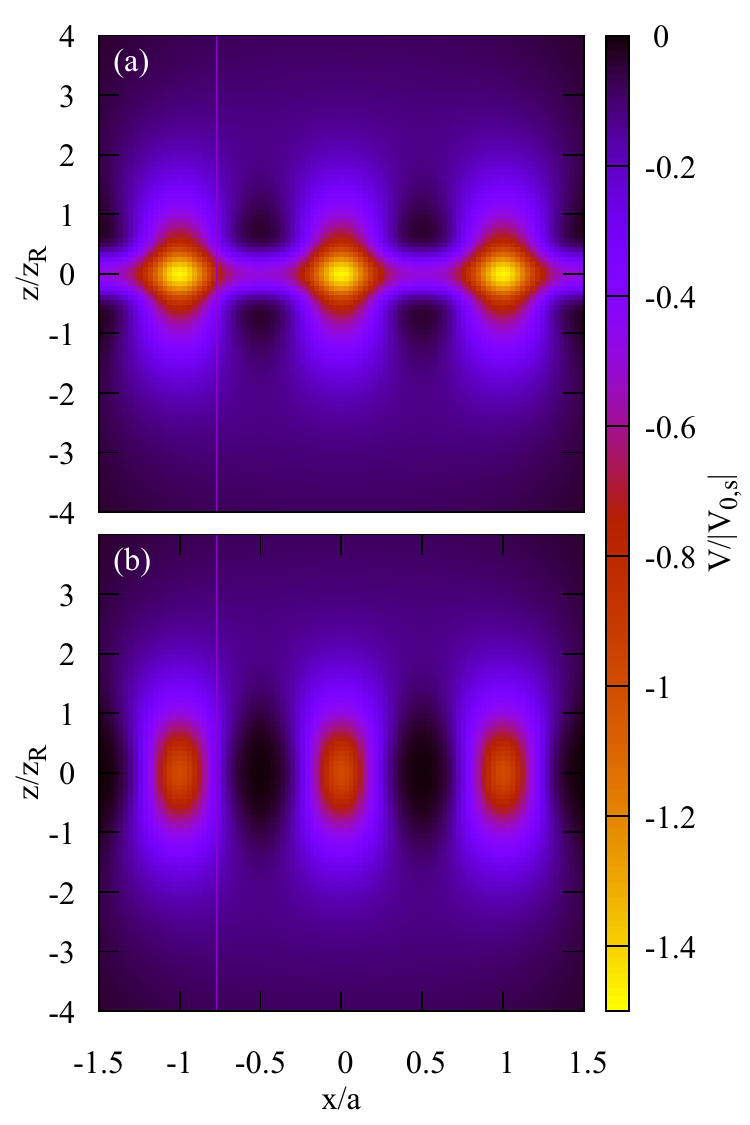}
\caption{(color online) (a) Superposition of spot and pancake potentials. The dark-gray (magenta) bar across the plot is the pancake, which also extends into the $y$-direction. The light-gray (yellow) peaks are the spot potential. (b) As (a), plotted without the pancake potential.}
\label{fig:potentialoverview}
\end{figure}

\begin{figure}
    \centering
    \includegraphics[width=85mm]{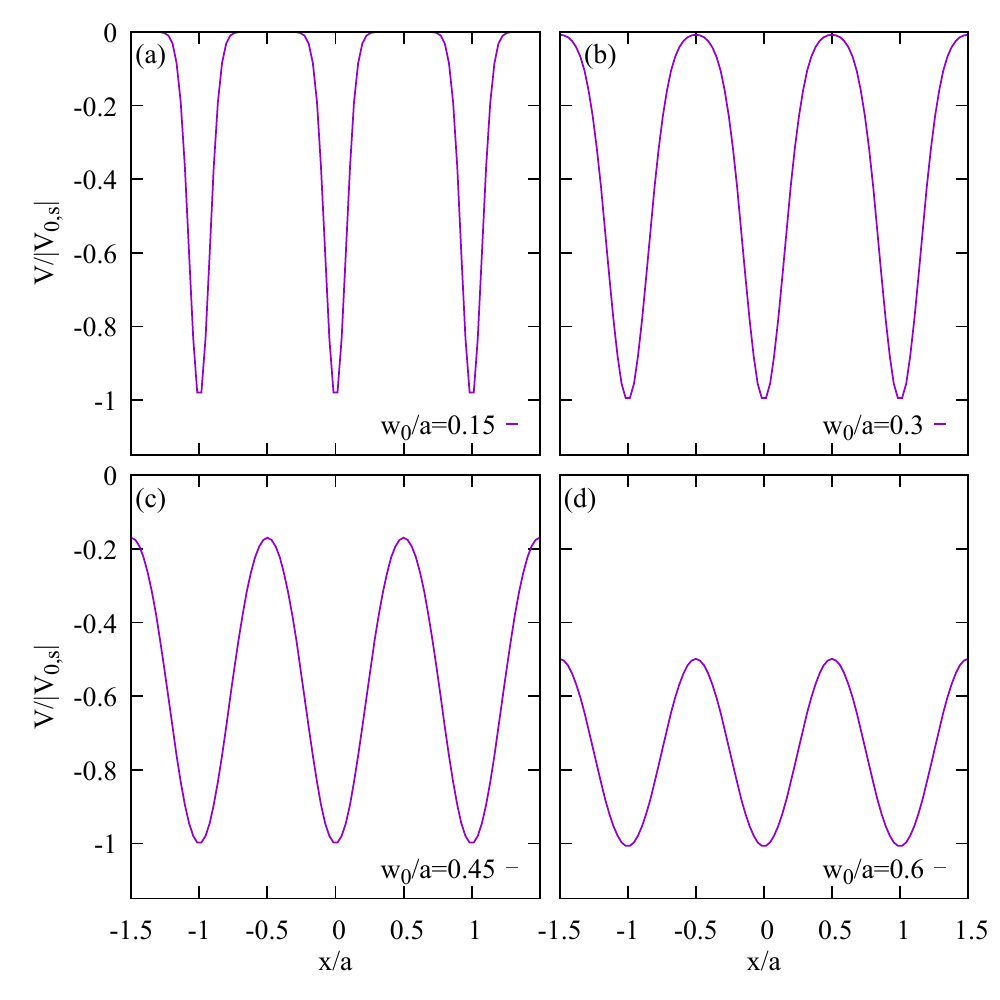}
    \caption{Effect of increasing beam waist on a periodic array of painted potentials. For small waist, spots have the form of an array of independent Gaussians. As the waist increases, they overlap to form sinusoids.}
    \label{fig:potentialxaxis}
\end{figure}

The form of this potential in the $xz$-plane is summarized in Fig. \ref{fig:potentialoverview}, and its form along the $x$-axis in Fig. \ref{fig:potentialxaxis}. In these figures, the spots are laid out on a regular lattice with intersite spacing, $a$. In Fig. \ref{fig:potentialoverview}(a), the optical pancake has been made unusually deep and narrow and can be seen as a bar across the image. Spots with very high potential can be seen close to the axis in Fig. \ref{fig:potentialoverview}(b). Well spaced spots act as individual Gaussians in Fig. \ref{fig:potentialxaxis}(a) and (b). As the spot waist becomes wider, the Gaussian spots overlap, and for $\wspot/a\gtrsim 0.45$ form a shallower sinusoidal potential that becomes flat when the distance between the spots is on the order of the waist (Fig. \ref{fig:potentialxaxis}(c) and (d)).

The potentials painted in such a setup are a distinct paradigm to optical lattices formed with counterpropagating beams. Lattices formed with counterpropagating beams have a sinusoidal form, with a uniform lattice with a simple basis. With painted potentials, spots have a Gaussian form, and each spot can be manipulated separately to the others, so translational symmetry can be broken, or a basis can be painted into the optical pancake.

\section{Hamiltonian}
\label{sec:deepwells}

In this section, we derive the strong-correlation Hamiltonian associated with painted potentials of the form in Eqn. \ref{eqn:totalpotential}. Since the argument of the exponential of the spot potentials depends on both $\rvecd$ and $z$, the Schr\"odinger equation is not separable, and we cannot make use of the mapping from a three-dimensional to a one-dimensional Schr\"odinger equation, such as the mapping to the one-dimensional Mathieu equation that is used to calculate $t$ and $U$ for sinusoidal potentials \cite{bloch2008a}. We note that in the limit that the spots making up painted potentials are close together the potential becomes approximately sinusoidal \cite{hague2012a}, however this is not generally true.

\subsection{Second quantization}
 
To allow for the possibility of complicated painted potentials including sites with different depths and lattices without translational symmetry, we use second quantized notation. This automatically accounts for the different particle densities on different sites and can be written down in real space so that systems without translational symmetry can be studied.
 
The second quantized interacting Hamiltonian has the form,
\begin{equation}
    \hat{H} = \hat{H}_{0} + \hat{H}_{\rm int}
\end{equation}
where $\hat{H}_{0}$ is the non-interacting part of the Hamiltonian and $\hat{H}_{\rm int}$ the interacting part.

The non-interacting Hamiltonian can be written as,
\begin{equation}
    \hat{H}_{0} = \int\drm^3\rvec \sum_{\sigma} \Psi^{\dagger}_{\sigma}(\rvec)\left[-\frac{\hbar^2}{2M}\nabla^2 + V(\rvec)\right]\Psi_{\sigma}(\rvec)
    \label{eqn:noninthamiltoniansq}
\end{equation}
where $\Psi^{\dagger}_{\sigma}(\rvec) = \sum_{\nvec}\Phi_{\nvec}(\rvec)c^{\dagger}_{\nvec,\sigma}$ is the appropriate field operator,
and we build the field from a basis of site-local wavefunctions (Wannier functions). The potential $V(\rvec)\equiv V(\rvecd,z)$ is defined in Eqn. \ref{eqn:totalpotential}. We note that the $\Phi_{\nvec}$ are not required to be periodically placed, nor does each site have to be equivalent. Each $\Phi_{\nvec}$ is centered about $\Rvec_{\nvec}$. In the following, we use the notation $\Phi_{\ksite}(\rvecd,z)=\Phi(\rvecd-\Rvecd_{\ksite},z)$, where the subscript $\ksite$ also indicates that the $\Phi$ correspond to the specific values of $\wspot$ and $\vzspot$ at site $\ksite$.

The interacting part of the Hamiltonian has the form,
\begin{align}
    \hat{H}_{\rm int} & = \iint\drm^3\rvec \drm^3\rvec'\sum_{\sigma,\sigma'}\frac{g_{\sigma,\sigma'}}{2}\Psi^{\dagger}_{\sigma}(\rvec)\Psi^{\dagger}_{\sigma'}(\rvec')\nonumber\\
    & \hspace{30mm}\times\delta(\rvec-\rvec') \Psi(\rvec)_{\sigma}\Psi(\rvec')_{\sigma'}\\
& = \int\drm^3\rvec \sum_{\sigma,\sigma'}\frac{g_{\sigma,\sigma'}}{2} \Psi^{\dagger}_{\sigma}(\rvec)\Psi^{\dagger}_{\sigma'}(\rvec)\Psi_{\sigma}(\rvec)\Psi_{\sigma'}(\rvec)
 \label{eqn:inthamiltoniansq}
\end{align}
We consider the non-interacting and interacting parts of the Hamiltonian in turn to calculate Hubbard parameters.

We examine the non-interacting Hamiltonian, Eqn. \ref{eqn:noninthamiltoniansq}, first. Expanding this equation, we find that,
\begin{equation}
    \hat{H}_{0} = \sum_{\nvec,\mvec,\sigma} c^{\dagger}_{\nvec,\sigma}c_{\mvec,\sigma} \langle\Phi_{\nvec}(\rvec)|-\frac{\hbar^2}{2M}\nabla^2 + V(\rvec)|\Phi_{\mvec}(\rvec)\rangle
\end{equation}
Thus, we obtain the following non-interacting Hamiltonian,
\begin{equation}
    \hat{H}_{0} = \sum_{\nvec}E_{\nvec}c^{\dagger}_{\nvec,\sigma}c_{\nvec,\sigma} + \sum_{\nvec\neq\mvec}t_{\rm \nvec\mvec} c^{\dagger}_{\nvec,\sigma}c_{\mvec,\sigma}.
    \label{eqn:noninthamiltonian}
\end{equation}
with
\begin{eqnarray}
   t_{\nvec\mvec}=\langle\Phi_{\nvec}|-\frac{\hbar^2}{2M}\nabla^2 + V(\rvec)|\Phi_{\mvec}\rangle.
    \label{eqn:hoppingfull}
\end{eqnarray}
and the local potential offset associated with individual lattice sites, %
%\begin{align}
%    -\frac{\hbar^2}{2M}\nabla^2 + V_{\rm spot,\mvec}(\rvec) & |\Phi_{\mvec}(\rvec)\rangle\nonumber\\ 
%    = E_{\mvec} & |\Phi_{\mvec}(\rvec)\rangle
%\end{align}
\begin{eqnarray}
   E_{\nvec}=\langle\Phi_{\nvec}|-\frac{\hbar^2}{2M}\nabla^2 + V(\rvec)|\Phi_{\nvec}\rangle.
    \label{eqn:energyoffset}
\end{eqnarray}
%\begin{equation}
%    -\frac{\hbar^2}{2M}\nabla^2 + V_{\rm site}(\rvec) |\Phi_{\mvec}(\rvec)\rangle = E_{\mvec} |\Phi_{\mvec}(\rvec)\rangle.
%\end{equation}
%

Expansion of the field operators in Eqn. \ref{eqn:inthamiltoniansq} shows that the interacting part of the many-body Hamiltonian (Eqn. \ref{eqn:inthamiltoniansq}) contains terms of the form,
\begin{equation}
    \hat{H}_{\rm int} = \sum_{\nvec\mvec\kvec\lvec\sigma\sigma'} c^{\dagger}_{\sigma\nvec}c^{\dagger}_{\sigma'\mvec}c_{\sigma\kvec}c_{\sigma'\lvec}U_{\nvec\mvec\kvec\lvec}
 \label{eqn:inthamiltonianhubbard}
\end{equation}
where
\begin{equation}
U_{\nvec\mvec\kvec\lvec\sigma\sigma'} = \frac{g_{\sigma,\sigma'}}{2}\int\drm^3\rvec  \Phi_{\nvec}(\rvec)\Phi_{\mvec}(\rvec)\Phi_{\lvec}(\rvec)\Phi_{\kvec}(\rvec)
\label{eqn:hubbarduraw}
\end{equation}

\subsection{Cold atoms in deep wells}

We proceed to find forms for $\Phi$ associated with the Gaussian spot potentials. To calculate $\Phi$ we require solutions to the Schr\"odinger equation for a single spot potential (Eqn. \ref{eqn:singlesplotpotenial}). It is not straightforward to obtain exact analytic expressions for this function. However, it is possible to obtain good analytic approximations to the wavefunction for a single spot, if the spot is very deep.

When the potentials are deep, atoms are expected to be highly localized. We expand Eqn. \ref{eqn:totalpotential} (the potential for both spots and the pancake) in $\rvecd$ and $z$ close to the center of a spot, giving,
\begin{eqnarray}
V(\rvecd,z)\approx V_{\text{HO}}(\rvecd,z) & = & -\left(\vzpan+\vzspot\right) +\frac{2\vzspot}{\wspot^{2}}|\rvecd|^2\nonumber\\
& & + \left(\frac{\vzspot}{z_{R}^{2}}+\frac{2\vzpan}{\wpan^{2}}\right)z^2. 
\end{eqnarray}
%

%where the subscript site stands for harmonic
%oscillator.

This potential has the form of a harmonic oscillator (HO). Comparison with $V_{\text{HO}}(\rvecd,z) = V_{\text{HO},0}+M\omega_{xy}^2 |\rvecd|^2/2+M\omega_{z}^2z^2/2$ leads to angular frequencies for the oscillator of,
\begin{align}
\omega_{xy} 
& = 2\sqrt{\frac{\vzspot}{M\wspot^2}},\label{eqn:omegaxy}\\
\omega_{z} 
& = \sqrt{\frac{2\vzspot}{M z_{R}^{2}}+\frac{4\vzpan}{M\wpan^2}}.\label{eqn:omegaz}
\end{align}
where $M$ is the atom mass and $V_{\text{HO},0}=-(\vzspot+\vzpan)$. Thus in the limit that the Gaussian spot potentials and pancake are deep, the
ground-state-spot orbital may be approximated with that of a harmonic oscillator,
\begin{equation}
\Phi\approx\Phi_{\text{HO}}(\rvecd,z)= \frac{M^{3/4}\omega_{xy}^{1/2}\omega_{z}^{1/4}}{\pi^{3/4}\hbar^{3/4}}e^{-\frac{M}{2\hbar}\left(\omega_{xy}|\rvecd|^{2}+\omega_{z}z^{2}\right)}
\end{equation}
(see e.g. \cite{ray}). We will use these wavefunctions as approximations to the Wannier functions to calculate the hopping integral and Hubbard $U$. As before, we use the notation $\Phi_{\text{HO},\lsite}(\rvecd,z)=\Phi_{\text{HO}}(\rvecd-\Rvecd_{\lsite},z)$, where the subscript $\lsite$ also indicates that the angular frequencies correspond to the spot potential at site $\lsite$. We note that similar approximations have been used to calculate Hubbard parameters for sinusoidal lattices \cite{jaksch1998, bloch2008a}.

The pancake potential does not contribute to the angular frequencies if $\vzpan/\wpan^2\ll\vzspot/\zr^2$. Since $\zr=\pi\wspot^2/\lambda_{\rm Las}$, and typically the spot waist is of the order, $\wspot\sim\lambda_{\rm Las}$, the condition becomes, $\vzpan/\wpan^2\ll\vzspot/\wspot^2$. Typically $\wpan\sim 500\,\mu\textrm{m}$ is three orders of magnitude larger than $\wspot\sim 500\,\textrm{nm}$. The pancake depth is of similar order, 1\,$\mu\mathrm{K}$ vs $500\,\textrm{nK}$. Thus the condition $\vzpan/\wpan^2\ll\vzspot/\wspot^2$ is well satisfied and we can neglect the pancake in the following.

The ground-state orbital has characteristic lengthscales $l_{xy}$ in the plane of the pancake and $l_{z}$ out of plane. The lengthscale within the plane is,
\begin{equation}
    l_{xy}=\sqrt{\frac{\hbar}{M\omega_{xy}}} = \left(\frac{\hbar\wspot}{2\sqrt{M\vzspot}} \right)^{1/2}.
\end{equation} 
If $\vzpan$ is small relative to $\vzspot$, then the lengthscale out of plane is,
\begin{equation}
    l_{z}=\sqrt{\frac{\hbar}{M\omega_{z}}} = \left(\frac{\hbar z_{R}}{\sqrt{2M \vzspot}} \right)^{1/2}.
\end{equation} 

We require that there are well localized bound states, such that the harmonic oscillator is a good approximation to the actual states.  For a state to be bound within the spot, we require that,
\begin{equation}
    l_{z}\ll z_{R}
    \label{eqn:conditionlz}
\end{equation}
and
\begin{equation}
    l_{xy}\ll \wspot.
    \label{eqn:conditionlxy}
\end{equation}
Substituting for $\omega_{xy}$ and $\omega_{z}$ from Eqns. \ref{eqn:omegaxy} and \ref{eqn:omegaz} (assuming that $\vzpan=0$ in the latter case), and rearranging, we see that,
\begin{equation}
    z_{R}\gg \frac{\hbar}{\sqrt{2M\vzspot}}
\end{equation}
and
\begin{equation}
    \wspot\gg \frac{\hbar}{2\sqrt{M\vzspot}}
\end{equation}
Up to a numerical factor, a similar expression can be derived by considering whether the lowest energy state of the harmonic oscillator is above the rim of the spot potential, $\hbar\omega_{z},\hbar\omega_{xy}\ll \vzspot$. 

\begin{figure}
    \centering
    \includegraphics[width=85mm]{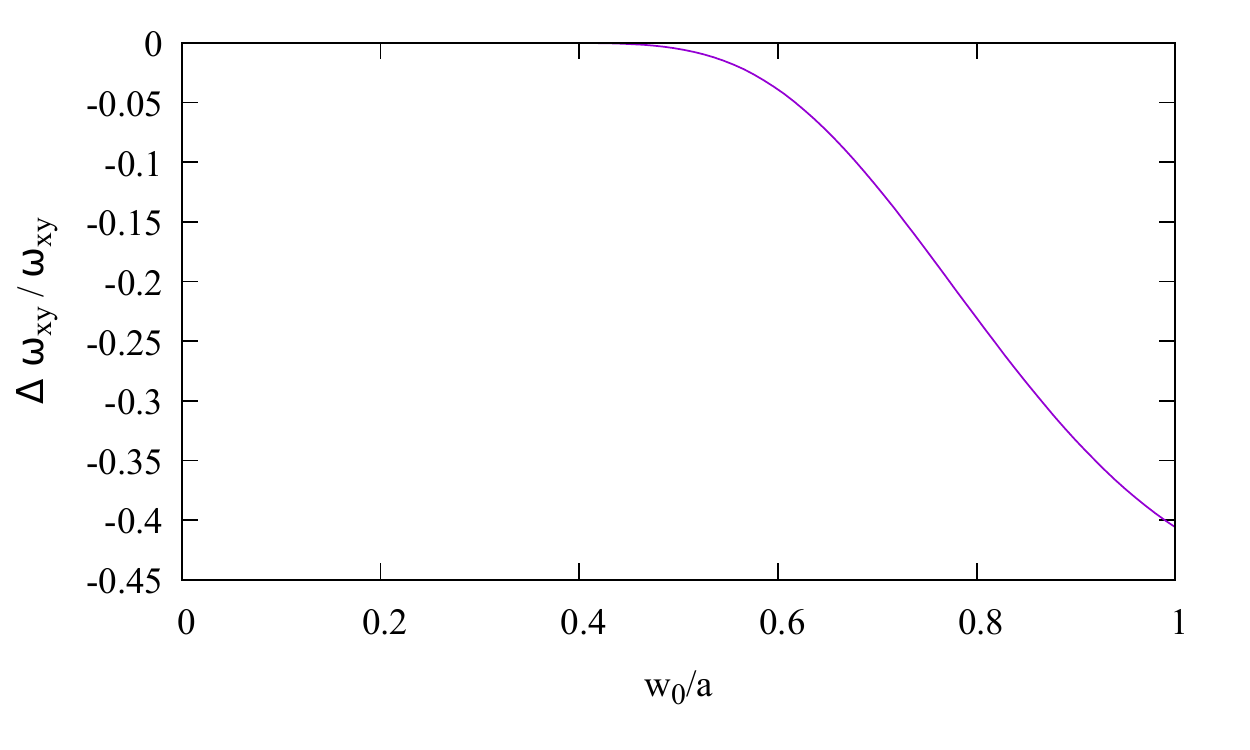}
    \caption{Correction to $\omega_{xy}$ given by Eqn. \ref{eqn:correction}. Corrections are negligible until $\wspot/a\sim 0.45$.}
    \label{fig:condition}
\end{figure}

This analysis requires that the spot potentials are sufficiently well separated that neighboring spots do not contribute to the harmonic oscillator frequency. If nearest-neighbor (nn) spots are included in the Taylor expansion, the second-order term for the potential in the plane becomes,
\begin{equation}
V_{\text{HO}}^{(nn)} = \frac{2\vzspot}{\wspot^2}\left[1+e^{-2a^2/\wspot^2}\left(1-\frac{4a^2}{\wspot^2}\right)\right]|\rvecd|^2
\end{equation}
where $a$ is the distance between the spots. Thus, the effect of neighboring spots on the harmonic oscillator frequency can be neglected if
\begin{equation}
    e^{-2a^2/\wspot^2}\left(1-\frac{4a^2}{\wspot^2}\right)\ll 1.
    \label{eqn:correction}
\end{equation}
This condition is plotted in Fig. \ref{fig:condition}. It can be seen that corrections to the harmonic oscillator frequency are negligible until $\wspot/a\sim 0.45$. This condition can be combined with Eqn. \ref{eqn:conditionlxy} to obtain,
\begin{equation}
    l_{xy}\ll \wspot \lesssim 0.45 a.
\end{equation}

\subsection{Hopping}

A goal here is to derive analytic expressions
for convenient estimation of the Hubbard parameters. We now apply the approximation of deep potentials to determining the hopping and Hubbard $U$. Replacing the Wannier functions, $\Phi$ with $\Phi_{\text{HO}}$ we obtain,
\begin{align}
    \hat{H}_{0} \approx & \sum_{\nvec}( K_{\nvec\nvec} + T_{\rm \nvec\nvec})c^{\dagger}_{\nvec,\sigma}c_{\nvec,\sigma}\\
    & + \sum_{\nvec\neq\mvec}c^{\dagger}_{\nvec,\sigma}c_{\mvec,\sigma}( K_{\nvec\mvec} + T_{\rm \nvec\mvec} )
\end{align}
with
\begin{align}
T_{\nvec\mvec} = & \langle \Phi_{{\text{HO}},\nvec}| \sum_{\lvec}V_{{\rm spot},\lvec}(\rvecd,z)|\Phi_{{\text{HO}},\mvec}\rangle.
\label{eqn:hoppingintegralsum}
\end{align}
(again noting the shorthand defined in Eqn. \ref{eqn:totalpotential} for spots centered about site $\Rvecd_{\lvec}$) and
\begin{equation}
K_{\nvec\mvec}=\langle\Phi_{{\text{HO}},\nvec}|-\frac{\hbar^2}{2M}\nabla^2|\Phi_{{\text{HO}},\mvec}\rangle
\end{equation}
Note that the sum in Eqn. \ref{eqn:hoppingintegralsum} is over all sites.

To make the calculation of $T_{\nvec\mvec}$ more amenable to analytic calculations we make a series of approximations. These take advantage of the exponentially decreasing tail of $\Phi_{\rm {\text{HO}}}(\rvec)$ on all axes and $V_{\rm spot,i}(\rvecd,z)$, within the $xy$-plane \footnote{This argument would not work with Coulomb lattice potentials, as they have long range tails, however there are no long range tails on spot potentials allowing truncation of the sum}. Thus, the sum in the $xy$-plane in Eqn. \ref{eqn:hoppingintegralsum} can be truncated to the spots located at sites at the beginning and end of the hop. This truncation is a good approximation since: (1) $V_{\rm spot}(\rvec)$ drops off exponentially in space and therefore the potentials associated with sites far from the initial and final sites in the hop are weak, and (2) the Gaussian wavefunctions associated with the harmonic oscillator tend to zero rapidly with distance, further reducing the contribution of the potentials from more distant lattice sites \footnote{We note that this argument would not work for an unscreened Coulomb potential in a traditional condensed matter tight-binding approximation.}. 

Following these considerations, the energy of the ground state within a single spot is expected to be a good approximation to the site local energy $E_{\nvec}$ in Eqn. \ref{eqn:noninthamiltonian}, i.e.
\begin{equation}
    E_{\nvec} \approx E_{{\text{HO}},\nvec} = -\vzspota+\hbar(2\omegaxya+\omegaza)/2
\end{equation}

We now turn our attention to the hopping. We break up $T_{\lsite\ksite}$ so that each term in the truncated potential sum is treated separately, writing $T_{\lsite\ksite}\approx T'_{\lsite\ksite}+T'_{\ksite\lsite}$. The integrals $T'_{\lsite\ksite}$ can be calculated from:
\begin{align}
T'_{\ksite\lsite} = & \langle \Phi_{{\text{HO}},\nvec}| V_{{\rm spot},\lsite}(\rvecd,z)|\Phi_{{\text{HO}},\mvec}\rangle\\ 
= & \iint \drm^{2}\rvecd \,\drm z V_{{\rm spot},\lsite}(\rvecd,z)\nonumber\\
   & \hspace{10mm}\times \Phi_{{\text{HO}},\ksite}^{*}(\rvecd,z) 
   \Phi_{{\text{HO}},\lsite}(\rvecd,z).\label{eqn:hoppingintegralapprox1}
\end{align}
We select the intersite vector relevant to this hopping term to lie along the $x$-axis such that $\avec = a\ivec$. Since the spot potentials may be different on sites $\lsite$ and $\ksite$ it is generally the case that $T'_{\lsite\ksite}\neq T'_{\ksite\lsite}$

We then note that the potential far from the $xy$-plane does not contribute strongly to the integral in Eqn. \ref{eqn:hoppingintegralapprox1}, especially if the potential is deep and the atoms are well confined to the $xy$-plane. Therefore, we Taylor expand $V_{\rm spot}(\rvecd,z)$ along the $z$-direction to second order. We shall call the approximate potential generated by this expansion $\tilde{V}(\rvecd,z)$; $z$-integrals for any of the terms that are generated in this way are straightforward to carry out. The Taylor expansion in $z$ is a good approximation if $l_{z}<~ z_{R}$, which will be true if the potential is sufficiently deep. The expanded potential has the form,
\begin{equation}
    \tilde{V}(\rvecd,z) =  -\vzspot\exp\left(-\frac{2|\rvecd|^2}{\wspot^2}\right)\left(1 - \frac{z^2}{z_{R}^2}\left(1-\frac{2|\rvecd|^2}{\wspot^2}\right)  \right).
\end{equation}
   
Once these simplifications have been made, $T'_{\ksite\lsite}$ has the following form:
\begin{align}
T'_{\ksite\lsite} \approx & \iint \drm^{2}\rvecd\, \drm z \tilde{V}_{\lsite}(\rvecd,z)\nonumber\\
   & \hspace{10mm}\times  \Phi_{{\text{HO}},\ksite}^{*}(\rvecd,z) 
   \Phi_{{\text{HO}},\lsite}(\rvecd,z).
   \label{eqn:hoppingintegralapprox2}
\end{align}

\begin{figure}
    \centering
    \includegraphics{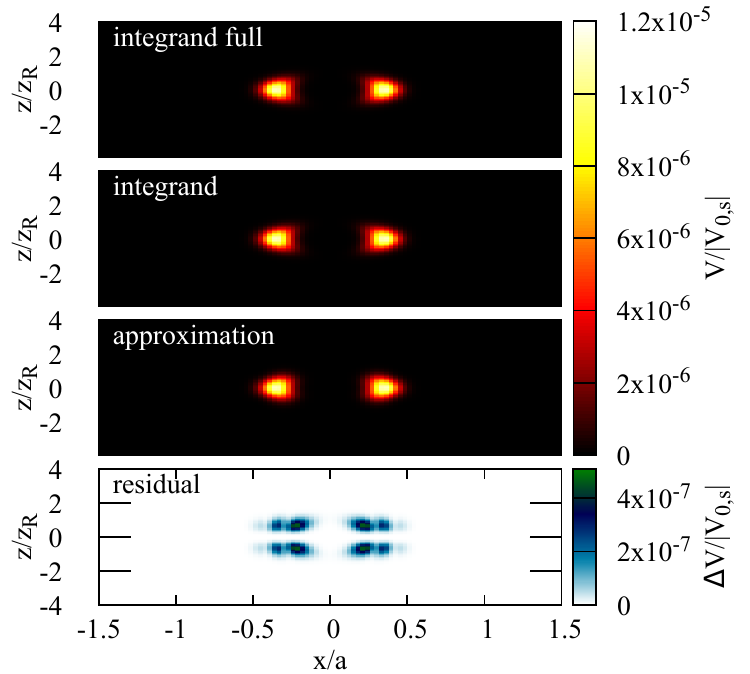}
    \caption{(color online) Comparison of integrand and approximate integrand for computing $T_{\lsite\ksite}$, when $\wspot/a=0.15$, $\vzspot=8\hbar^2/2Ma^2$, $\zr=a$. Integrand corresponding to: Eqn. \ref{eqn:hoppingintegralsum} (integrand full); $T'_{\lsite\ksite}+T'_{\ksite\lsite}$ with lattice potential only at initial and final hopping sites, consistent with Eqn. \ref{eqn:hoppingintegralapprox1} (integrand); $T'_{\lsite\ksite}+T'_{\ksite\lsite}$ with the approximations in Eqn. \ref{eqn:hoppingintegralapprox2} (approximation). The bottom panel shows the difference between the full and approximate integrands, which is small (residual).}
    \label{fig:integrand0x15}
\end{figure}

\begin{figure}
    \centering
    \includegraphics{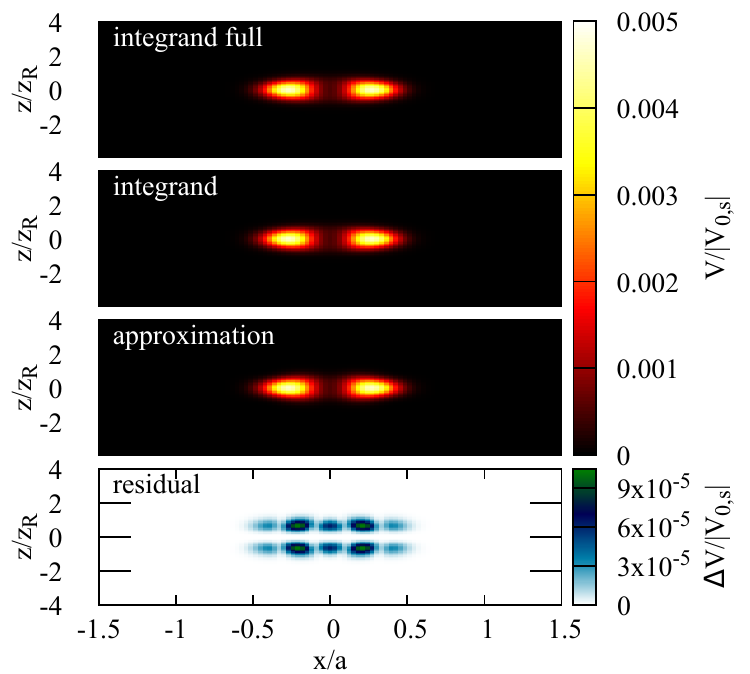}
    \caption{(color online) As Fig. \ref{fig:integrand0x15}, with $\wspot/a=0.3$, $\vzspot=8\hbar^2/2Ma^2$, $\zr=a$. The residual between the full and approximate integrands is tiny.}
    \label{fig:integrand0x3}
\end{figure}

To demonstrate the quality of the approximation, we plot the integrands of Eqns. \ref{eqn:hoppingintegralsum} and \ref{eqn:hoppingintegralapprox2} within the $xz$-plane in Figs. \ref{fig:integrand0x15} and \ref{fig:integrand0x3} for two different values of $\wspot/a$. For $\wspot/a=0.15$ (Fig. \ref{fig:integrand0x15}) the residual between the two integrands is not visible at the resolution of the color scale. When $\wspot/a=0.3$ (Fig. \ref{fig:integrand0x3}) a tiny difference can be made out. 

Thus, the hopping term has been rewritten in terms of standard Gaussian integrals. Such integrals have been studied extensively (see e.g. \cite{gill1994a}). We proceed by integrating in the order $z$-axis, $y$-axis, and finally $x$-axis. The resulting expression is,
\begin{widetext}
    \begin{align}\label{eqn:potentialcontributiontohopping}
    T'_{\ksite\lsite}
    & =-\frac{2 M^{1/2} \vzspota^{3/8} \vzspotb^{11/8} \wspota^{1/2} \wspotb^{3/2} \zra^{1/4}  
    \exp \left(-\frac{a^2 \sqrt{M \vzspota} \left(2 \hbar+\wspotb \sqrt{M \vzspotb}\right)}{\hbar \left(2 \hbar \wspota+\wspotb \sqrt{M}W\right)}\right)}{\zrb^{3/4} Z^{3/2} \left(2
   \hbar \wspota+\wspotb \sqrt{M}W\right)^3}\\
    & \hspace{8mm}\times 
   \Bigg[\hbar \wspotb\zra\bigg(\wspotb\sqrt{M} \Big(2 a^2 \vzspota
   - W^2\big)-2\hbar\wspota W\bigg)+\hbar\wspota\zrb Z\sqrt{32} \bigg(\wspotb W\sqrt{M}+\hbar\wspota\bigg)+\sqrt{2} M \wspotb^2
   \zrb W^2 Z\Bigg]\nonumber
   \end{align}
where $\vzspota$ and $\vzspotb$ denote the spot depth $\vzspot$ on sites $\ksite$ and $\lsite$, respectively; $\wspota$ and $\wspotb$ denote the spot waist $\wspot$ on sites $\ksite$ and $\lsite$, respectively; and $\zra$ and $\zrb$ denote the Rayleigh length on sites $\ksite$ and $\lsite$, respectively. We also define, $Z=(\sqrt{\vzspota} \zrb+\sqrt{\vzspotb} \zra)$ and $W=(\sqrt{\vzspota} \wspotb+\sqrt{\vzspotb} \wspota)$. 

The overlap integral of the kinetic energy operator is found to be,
  \begin{align}
  K_{\ksite\lsite}=& \frac{2(\vzspota \vzspotb)^{7/8} (\zra \zrb)^{1/4}\sqrt{\wspota\wspotb} }{M^{1/2}  W^3 Z^{3/2}}
   \Bigg(\hbar\left(W^2+ZW\sqrt{8}\right)- a^2 Z\sqrt{8M\vzspotb\vzspota}\Bigg)e^{-\frac{a^2
   \sqrt{M\vzspota \vzspotb}}{\hbar W}}
   \label{eqn:kecontributiontohopping}
   \end{align}

   From these, the hopping is constructed as,
   \begin{equation}
       t_{\ksite\lsite} = -(K_{\ksite\lsite}+T'_{\ksite\lsite}+T'_{\lsite\ksite}).
   \end{equation}

We can then determine the large $\vzspot$ (deep potential) behavior of these expressions, noting that $Z$ and $W$ are proportional to $\sqrt{\vzspot}$, the third term in Eqn. \ref{eqn:potentialcontributiontohopping} and second term of Eqn. \ref{eqn:kecontributiontohopping} dominate at large $\vzspot$. Therefore, the hopping is,
\begin{equation}
t_{{\rm deep},\ksite\lsite}=\frac{2^{3/2}\sqrt{\wspota\wspotb}(\vzspota\vzspotb)^{3/8} \sqrt[4]{\zra \zrb} e^{-\frac{a^2 \sqrt{M \vzspota\vzspotb}}{\hbar W}}}{ W^3
   Z^{1/2}} \Bigg(2 a^2  \vzspota\vzspotb+(\vzspota+\vzspotb) W^2\Bigg)
      \label{eqn:hopping_strong_coupling}
\end{equation}
   \end{widetext}
This is the key result of this paper. The hopping is very important because it sets the timescales of the quantum simulator, and is less easily tuned than the Hubbard $U$. The hopping is sensitive to the width, depth and the Rayleigh length associated with spots at the start and end of the hop.

In the event that sites $\ksite$ and $\lsite$ have the same depth and width, the hopping simplifies further, such that,
\begin{equation}
t_{\rm deep} = \left(2+\frac{a^{2}}{2\wspot^{2}}\right)\vzspot \exp\left(-\frac{a^2 \sqrt{M\vzspot
   }}{2 \hbar \wspot}\right)
   \label{eqn:hopping_strong_coupling_equal}
   \end{equation}
We note that this expression does not depend on $\zr (= \pi\wspot^2/\lambda_{\rm Las})$. Eqn. \ref{eqn:hopping_strong_coupling_equal} has a similar structure to hopping in simple sinusoidal lattices \cite{bloch2008a}. A semi-analytic expression based upon hopping in sinusoidal lattices was used to fit to the hopping by Wall {\it et al.} \cite{wall2015}. The key difference here is that our expression shows the dependence of hopping on the key optical lattice parameters, $\wspot$ and $\vzspot$, whereas the fit in Ref. \cite{wall2015} is for a single set of parameters. Furthermore, Eqn. \ref{eqn:hopping_strong_coupling} allows for cases where different sites represent different atom types in a material, which therefore have different $\wspot, \vzspot$ and $\zr$ to represent different nuclear potentials. In contrast to similar expressions for sinusoidal lattices (see e.g. Ref. \cite{bloch2008a}), the hopping depends on both lattice spacing and spot properties. We are not aware of any expressions similar to Eqn. \ref{eqn:hopping_strong_coupling} in the literature.

%This final expression can be cast in the following form for comparison with the single set of fit parameters obained by fitting to numerics in Wall et al. \cite{wall2015},
%\begin{equation}
%    t_{\rm deep}/E_{R}=A(V/E_{R})^B \exp(-C\sqrt{V/E_R})
%\end{equation}
%where $E_{R}=(h/a)^2/2M$, $A = 2+a^2/2\wspot^2$, $B=1$ and $C=a\pi/\wspot\sqrt{2}$. These are consistent with the values obtained to a fit to numerics by Wall et al. \cite{wall2015}: $A=2.563$, $B=1.217$ and $C=2.281$.

% $\sqrt{E_{R}}=2\pi \hbar/a\sqrt{2M}$.

\subsection{Hubbard $U$}
\label{subsec-HubbardU}

The Hubbard $U$ is a critical part of the quantum simulator, since it controls the level of interaction between atoms, thus making the quantum simulation non-trivial. The most interesting regime of Hubbard models occurs when the magnitude of $U\sim t$, and so control over $U$ is extremely important. In this section, we derive an approximation for the Hubbard $U$ in the case that the spot potentials are deep.

Starting from Eqn. \ref{eqn:hubbarduraw}, we note that, since atoms are well localized to optical lattice sites, this integral is largest if the site indices are shared. The biggest of these are expected to be the Hubbard coefficients, $U_{\nvec\mvec}$, which can be determined using the expression\cite{bloch2008a}.
\begin{equation}
U_{\nvec\mvec,\rm Fesh}=g\int \Phi^{2}_{\nvec}(\rvecd,z)\Phi^{2}_{\mvec}(\rvecd,z)\drm^2\rvecd\drm z
\end{equation}
note that the factor $1/2$ in Eqn. \ref{eqn:hubbarduraw}
 canceled due to counting of pairwise interactions in both directions. Again note that in our compact notation, the subscripts $\lsite$ and $\ksite$ are centered about $\Rvecd_{\lsite}$ and $\Rvecd_{\ksite}$ respectively. 
Here we have combined the Hubbard terms into a single function, $U_{\ksite\lsite,\rm Fesh}$. As before, integrals within and perpendicular to the pancake are separated due to the asymmetry of the potential. We shall refer to $g=4\pi\hbar^2 a_{s}/2M$ as the interaction coupling constant, which represents the magnitude of the interaction mediated by the Feshbach resonance; $a_{s}=a_{\rm bg}(1-\Delta B/(B-B_{0}))$ is the scattering length in the vicinity of the Feshbach resonance (for fermionic, $^6$Li, the s-wave scattering length is $a_{\rm bg}\approx 2.9\,\mathrm{nm}$ and for fermionic $^{40}\mathrm{K}$, $a_{\rm bg}\approx 5.5\;\mathrm{nm}$); $B$ is magnetic field; and $\Delta B$ is the width of the Feshbach resonance. This expression allows for the possibility that different sites have different depths and waists.

We make the approximation,
\begin{equation}
U_{\ksite\lsite,\rm Fesh}\approx g\int \Phi_{{\text{HO}},\ksite}^{2}(\rvecd,z)\Phi_{{\text{HO}},\lsite}^{2}(\rvecd,z)\drm^2\rvecd\drm z
\end{equation}

This integral can be evaluated to obtain the expression,
%
%\begin{equation}
%  U^{\rm({\text{HO}})}_{\ksite\lsite,\rm Fesh} = g\left(\frac{M}{2\pi\hbar}\right)^{\frac{3}{2}} \bar{\omega}_{xy} \sqrt{\bar{\omega}_{z}} e^{-\frac{M \bar{\omega}_{xy} |\Rvecd_{\ksite\lsite}|^2}{2 \hbar}}
%\end{equation}
\begin{equation}
  U^{\rm({\text{HO}})}_{\ksite\lsite,\rm Fesh} = g \frac{\left(M \vzspota\vzspotb\right)^{3/4} 2^{5/4}}{W Z^{1/2} \pi^{3/2} \hbar^{3/2}} e^{-2\frac{\sqrt{M\vzspota\vzspotb}  |\Rvecd_{\ksite\lsite}|^2}{W\hbar}}.
\end{equation}
%
%where $\bar{\omega}_{z}=\omegaza\omegazb/(\omegaza+\omegazb)$ and $\bar{\omega}_{xy}=\omegaxya\omegaxyb/(\omegaxya+\omegaxyb)$. [JPH: MAKE NOTE THAT THIS IS FAIRLY STANDARD EXPRESSION]

Intersite $U$ will be small unless wavefunctions overlap strongly between sites. Overlap of wavefunctions will only occur if sites are very close together (within the lengthscale $l_{xy}$). Normally, the intersite Hubbard $U$ will be negligible and it will be sufficient to consider the onsite Hubbard $U$, where $\Rvecd_{\ksite\lsite}=0$. Dressed Rydberg atoms could be used to generate large intersite interactions.

The onsite Hubbard $U$ is,
%
%\begin{equation}
%  U^{\rm ({\text{HO}})}_{\ksite\ksite} = g\left(\frac{M}{2\pi\hbar}\right)^{\frac{3}{2}} \bar{\omega}_{xy} \sqrt{\bar{\omega}_{z}}.
%  \end{equation}

%Neglecting the effect of the optical pancake and substituting the expressions for $\bar{\omega}$ leads to:
%
\begin{equation}
  U^{\rm({\text{HO}})}_{\ksite\ksite,\rm Fesh} = g \frac{M^{3/4} \vzspota^{3/4}}{2^{1/4} \wspota \zra^{1/2} \pi^{3/2} \hbar^{3/2}},
\label{eqn:onsiteufinal}
\end{equation}
or alternatively,
\begin{equation}
    \frac{U_{\ksite\ksite}^{({\text{HO}})}}{g} = \frac{\vzspota^{3/4}M^{3/4}\lambda_{\rm Las}^{1/2}}{2^{1/4}\pi^2\hbar^{3/2}\wspota^2}
%  \label{eqn:onsiteufinal}
\end{equation}
where the expression depends on $z_{R}$ and therefore $\lambda_{\rm Las}$.

This is the second key result in the paper. There is a similar $V^{3/4}$ functional form to the Hubbard $U$ in sinusoidal lattices, and the $V^{3/4}$ dependence has been noted by Wall {\it et al.} \cite{wall2015}, without detailed prefactors giving the dependence on $\wspot$ and $\zr$. The absolute magnitude of the Hubbard $U$ is very important for interaction, but is less important than the hopping in quantum simulator design, since $U$ is easily tuned by varying the magnetic field through the Feshbach resonance. However, the relative sizes of the Hubbard $U$ interactions on different sites is important for quantum simulator design, and so knowing the dependence of Eqn. \ref{eqn:onsiteufinal} on $\wspot$ and $\zr$ is essential.

\subsection{Tuneability}

Equations \ref{eqn:hopping_strong_coupling_equal} and \ref{eqn:onsiteufinal} show how painted potentials can be used
for quantum simulations of strongly correlated Hamiltonians. There are
several ways of tuning the relative interaction strength: 
\begin{enumerate}
    \item In the
same way as Mott--Hubbard simulators using sinusoidal potentials by changing $\vzspot$ or
tuning the Feshbach resonance.
\item By modifying the intersite distance
$a$, which can be tuned without changing the laser frequency.
\item By modifying the spot
width of the optical tweezers, $\wspot$. 
\item By modifying the laser frequency, $\lambda_{\rm Las}$ (although this is difficult in practice).
\end{enumerate}

\begin{figure*}
    \centering
    \includegraphics[width=140mm]{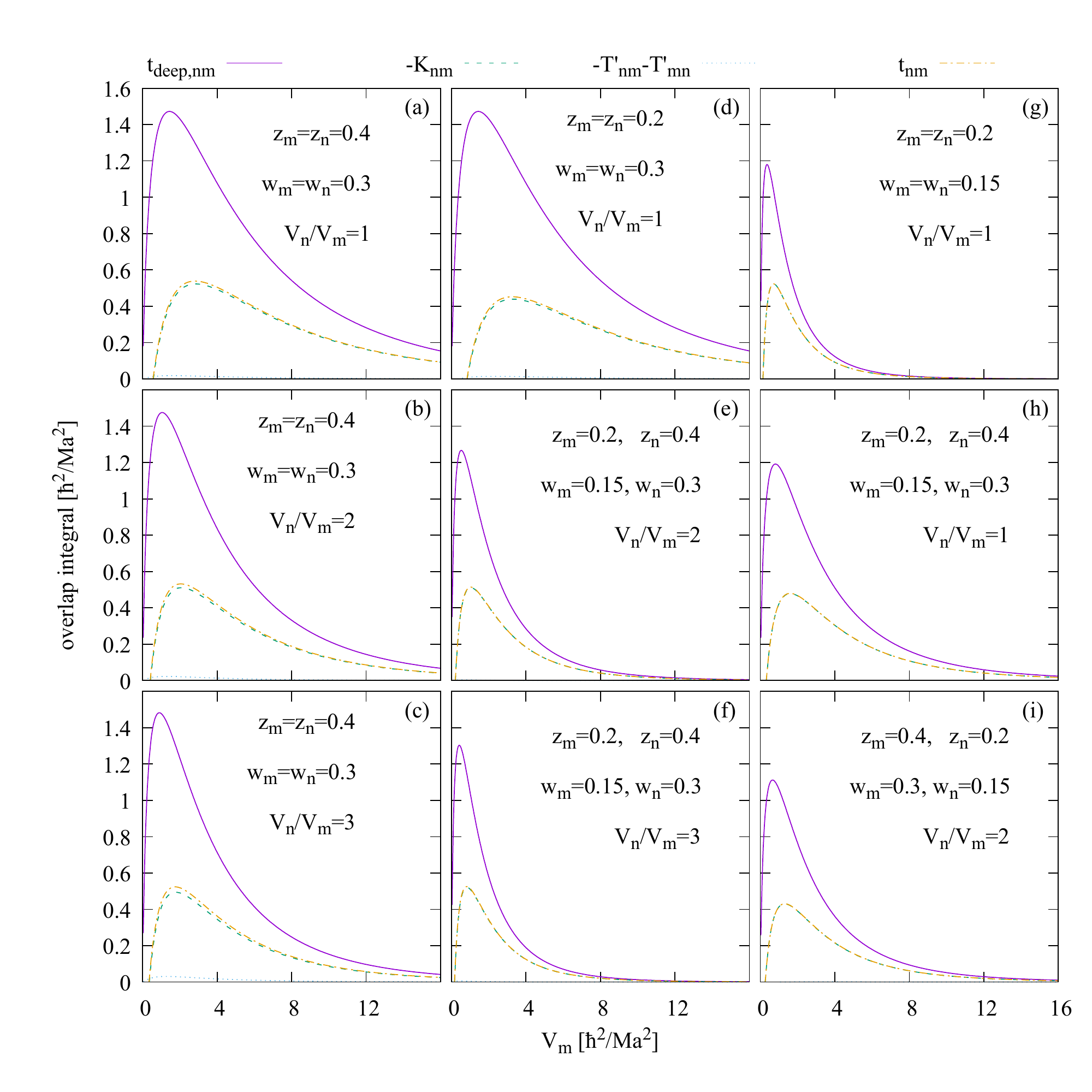}
    \caption{Variation of the hopping, $t$, with $\wspot$, $\zr$, and $\vzspot$. For the calculations we set $M=1, a=1, \hbar=1$ leading to results with the units $\hbar^2/2Ma^2$. A wide range of hopping parameters can be achieved by modifying the spot properties.} 
    \label{fig:hopping}
\end{figure*}

\begin{figure}
    \centering
    \includegraphics[width=90mm]{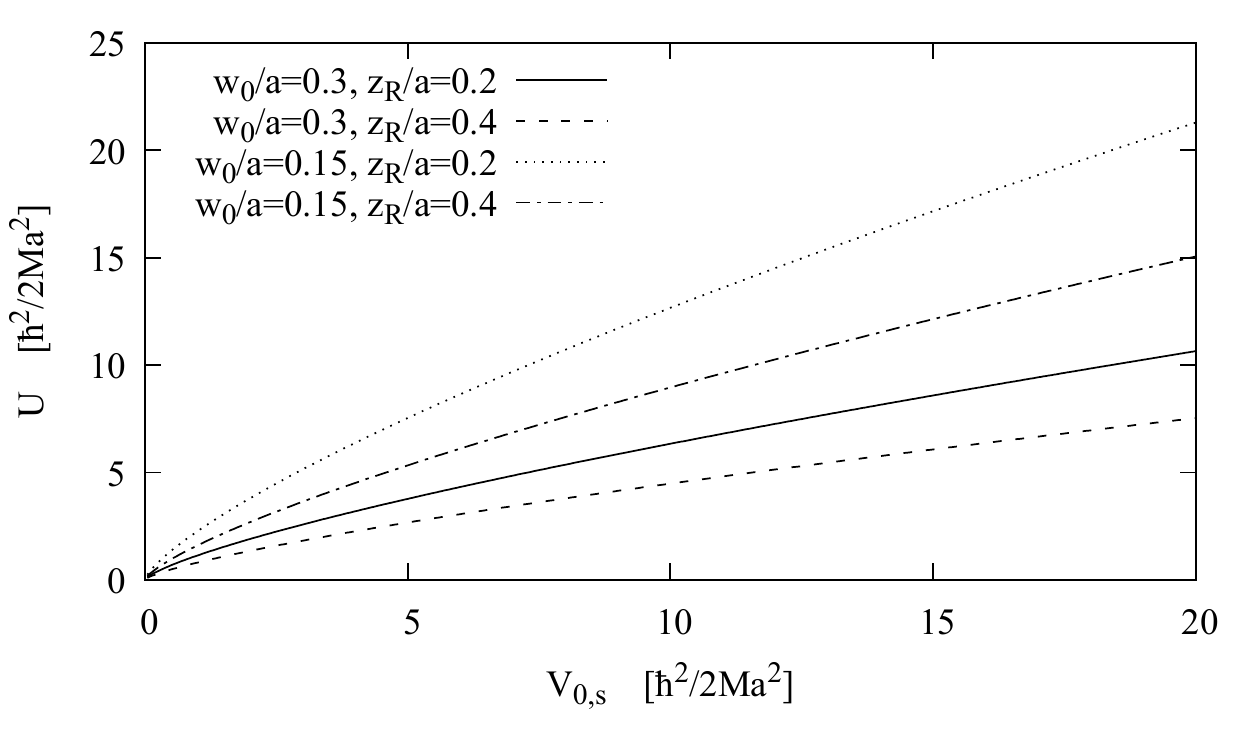}
    \caption{Variation of the Hubbard $U$ with $\wspot/a$ and $z_{R}/a$. For the calculations we set $g=1$, $M=1, a=1, \hbar=1$ leading to results with the units $\hbar^2/2Ma^2$. For $g=1$, the Hubbard $U$ has a magnitude of a few $t$. $U$ scales linearly with $g$, so is easily tuned by changing the magnetic field associated with the Feshbach resonance.}
    \label{fig:hubu}
\end{figure}

We show examples of this tunability in Figs. \ref{fig:hopping} and \ref{fig:hubu}. To set the energy scale, we have set $\hbar=M=a=1$, so units of the hopping and Hubbard $U$ are $\hbar^{2}/Ma^2$. 

The effect of different spot depths and widths on initial and final sites of a hop is shown in Fig. \ref{fig:hopping}. Comparison of panels (a) and (d) shows that hopping is not strongly dependent on $z_{R}$. In contrast, comparison of panels (d) and (g) shows how halving $\wspot$ leads to a significant decrease in the hopping rate. Panels (a), (b), and (c) show the effect of increasing the depth of one spot relative to another when spot width is unchanged (which effectively scales the $x$ axis). Panels (e), (h), and (i) show the counter-intuitive result that modifying the ratios of spot depths can lead to a small increase in hopping when other spot parameters are not identical. While spots have to be very deep to recover large $\vzspot$ behavior to high accuracy, the large $\vzspot$ behavior can be used as an estimate of the full expression for hopping. 

With $g=1$ the Hubbard $U$ has the magnitude of a few $t$, as can be seen in Fig. \ref{fig:hubu}. The Hubbard $U$ increases with $\vzspot$; $g$ is highly tunable, and can be used to decrease or increase $U$ as required to access regions of experimental interest. The Hubbard $U$ depends on $z_{R}$ and $\wspot$, so these can also be used to tune the interaction.

\section{Implementation}
\label{sec:implementation}

\begin{figure}
    \centering
    \includegraphics[width=85mm]{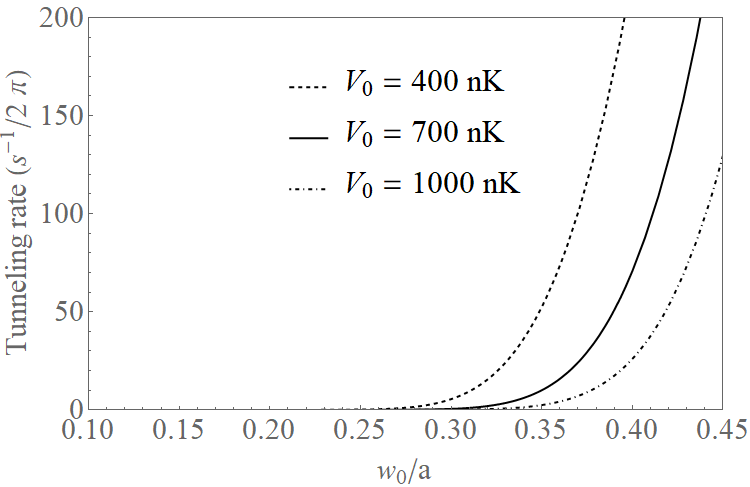}
    \caption{Hopping vs $\wspot/a$ calculated using Eqn.~(\ref{eqn:hopping_strong_coupling}), for an array of identical Gaussian traps of spot size $\wspot = 0.7 \;\mu\mathrm{m}$, and trap depths $\vzspot = 400 \;\mathrm{nK}$,$\vzspot = 700 \;\mathrm{nK}$ and $\vzspot = 1000 \;\mathrm{nK}$.
    }
    \label{fig:hopping_traps}
\end{figure}

\begin{table*}
  \caption{Summary of the quantum simulator and correspondence with condensed matter systems. Items marked with a dash are not possible using optical lattices formed with purely counterpropagating laser beams.}
  \begin{tabular}{|c|c|c|c|}
  \hline
 &  \multicolumn{2}{c|}{\bf quantum simulator} & {\bf condensed matter} \\
    \hline
  \textbf{fermion} & \multicolumn{2}{c|}{fermionic atom, e.g. $^{40}$K / $^{6}Li$} & electron \\
  \hline
  \textbf{lattice potential}; \textit{\textbf{origin:}} & painted spot potential & counterpropagating laser beams & nuclear potential\\
 \textit{\textbf{form:}}     & (Gaussian) & (sinusoidal) & ($1/r$) \\
      \hline
\textbf{Hubbard $U$};
\textit{\textbf{origin:}} & Feshbach resonance & Feshbach resonance  & Coulomb repulsion\\
\textit{\textbf{site scale tuning:}} & individual spot depth and width & - & impurities \\
\textit{\textbf{lattice scale tuning:}} & global spot depth and width & lattice depth & pressure\\
\hline
\textbf{hopping, $t$}; \textit{\textbf{origin:}} &    QM tunneling &    QM tunneling & QM tunneling\\
\textit{\textbf{site scale tuning:}} & individual spot depth and width & - & impurities\\
\textit{\textbf{lattice scale tuning:}} & spot spacing &  lattice depth & pressure \\
\hline
    \end{tabular}
    \label{tab:qs}
  \end{table*}

In this section, we estimate parameters for an experimental quantum simulator. We select $^{6}$Li as a suitable fermionic atom which has been widely used in cold-atom experiments. Details of the Feshbach resonances in this atomic system can be found in Ref. \cite{Schunck2005}.
% {\color{red} [CM: changed to 6LI, since 6Li is the Fermion with 9 particles]}

To control the quantum simulator, it is possible to tune $\vzspot$, $\wspot$, and $a$. The values of $\wpan$ are not used to tune parameters.  We do not count $\lambda_{\rm Las}$ as a (convenient) way of tuning the properties of the optical lattice. While it can be varied, its value is limited by the transitions that form the dipole potential according to Eqn. \ref{eqn:vdipfromtransition}, which in turn are dictated by the type of atoms we are using.

The strong D1 and D2 transitions in $^{6}$Li have a saturation intensity $I_\mathrm{Sat}=25.4\, \mathrm{W}\,\mathrm{m}^{-2}$, linewidth $\Gamma=2\pi\times 5.87\times 10^6\;\mathrm{s}^{-1}$, and closely separated wavelengths of $\lambda_\mathrm{D1}=670.979$ nm and $\lambda_\mathrm{D2}=670.977$ nm respectively. With these parameters, an 852 nm laser (red detuned from the D1 and D2 transitions) focused to a spot size of $0.7\,\mu\mathrm{m}$ will achieve a trap depth of $110\,\mathrm{nK} \,\mu\mathrm{W}^{-1}$. Thus typical $\sim 100$mW lasers can produce thousands of traps with a depth 400--1000 nK. 

Such trapping potentials are demanding. In this work, we assume that the spot sizes are produced by low-abberation (i.e. spherical abberations),  diffraction limited set ups as pioneered in Ref. \cite{sortais2007}, where an 852 nm laser was focused to $0.9 \,\mu\mathrm{m}$ using a high quality aspheric lens with a numerical aperture of 0.5. More recent work improved upon these conditions by moving the lens much closer to the atoms and using a considerably higher numerical aperture (NA) \cite{Bakr2009}.

The hopping (tunneling) rate between sites is calculated in Fig. \ref{fig:hopping_traps} for three different trap depths (labeled with their equivalent temperatures), with the hopping rate expressed in Hz. The exponential terms in the hopping lead to rapid increases in $t$ between $\wspot/a=0.25$ and $\wspot/a=0.3$ depending on the depth of the spot potentials. Hopping rates of a few hundred Hz are consistent with those of optical lattices formed from sinusoidal potentials.

Now we turn to a discussion of sources of decoherence in this system. The trapping lasers are a source of heating for experiments such as these. The trapped atoms scatter photons from the far-detuned trap lasers at a rate $R_\mathrm{S} = \Gamma(\Omega/2\Delta)^2 $, where, in terms of the laser Intensity $I_\mathrm{Las}$, the Rabi frequency  squared is expressed as $\Omega^2 = \Gamma^2 I_\mathrm{Las}(\mathbf{r})/2I_\mathrm{Sat}$. On average, each scattering event is associated with a recoil energy $E_\mathrm{R}=(h/a)^2/2M$, so the heating rate can be estimated as $\dot{E}_\mathrm{Heat} = E_\mathrm{R}R_\mathrm{S}$. For the parameters typical of this work, we have $E_\mathrm{R} = 1.59/a^2 \; \mu\mathrm{K}$ (where $a$ has units of $\mu\mathrm{m})$. The heating is due to the cumulative effect of the lattice and pancake potentials. We assume that a suitable pancake would be realized using a $852\;\mathrm{nm}$ laser focused to make a pancake shape with beam waists of $500\,\mu\mathrm{m}$ and $10\,\mu\mathrm{m}$. With these parameters, a 100 mW laser will create an optical pancake of depth $1.1\;\mu\mathrm{K}$ and $\omega_{c,z}\sim 2\pi\times 1.2\; \mathrm{kHz}$; when $a=\wspot/0.35 = 2\,\mu\mathrm{m}$, we estimate the heating rate from the pancake beam is $ 3.52 \,\mathrm{nK}\;\mathrm{s}^{-1}$. For lattice potential with spot size $\wspot = 0.7 \;\mu\mathrm{m}$ and spot depth $\vzspot/k_{B}=750\mathrm{nK}$, the heating rate is  $ 2.42 \,\mathrm{nK}\;\mathrm{s}^{-1}$.

Finally, we comment on another potential source of decoherence, which is the possibility of dissociation caused by tuning the Hubbard $U$ using Feshbach resonances as outlined in subsection~\ref{subsec-HubbardU}. Dissociation occurs when the scattering length is about the same size as the mean inter-particle spacing.
In general, it is desirable to achieve $U\sim t$ and in this paper, we have found that $t$ is of order 10--100 Hz. For a cold gas at a temperature of $100 \,\mathrm{nK}$, and lattice spot-size $\wspot = 0.7\;\mu\mathrm{m}$, a doubly occupied site has $U\sim 90\;\mathrm{Hz}$, and so the degree of tuning required to achieve $U=t$ would lead to a maximum $a_\mathrm{s}\sim 10 a_\mathrm{bg} = 0.024\;\mu\mathrm{m}$, an order of magnitude smaller than the interparticle spacing which is of order $0.3\;\mu\mathrm{m}$. $a_{\rm bg}$ is defined in Sec. \ref{subsec-HubbardU}.

We also note that it is possible to simulate Bose--Hubbard models by changing $^{6}$Li for low-mass bosonic atoms (e.g. $^{7}$Li or even metastable helium \cite{Keller2014}).

\begin{figure}
    \centering
    \includegraphics[width=80mm]{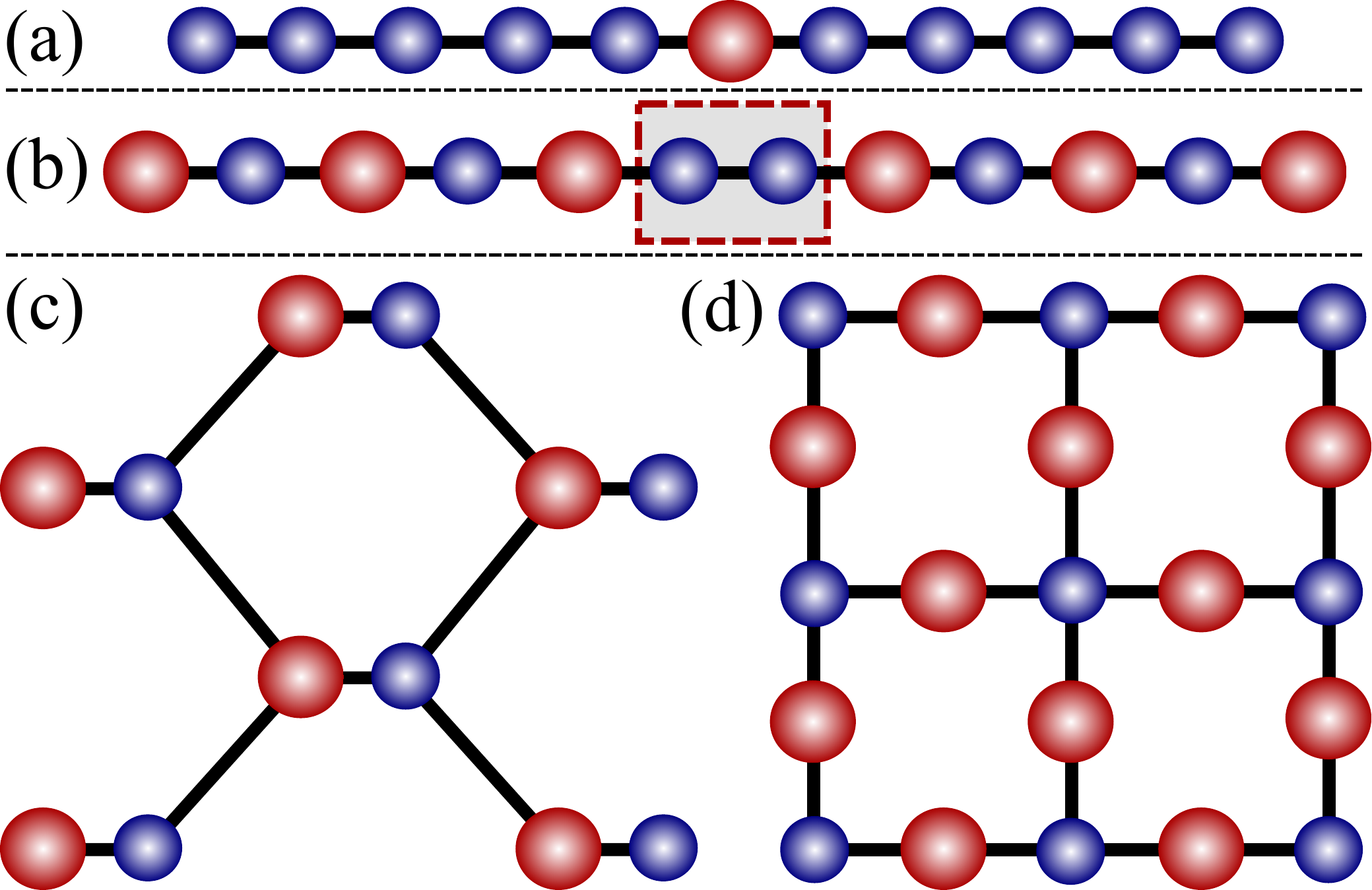}
    \caption{(color online) Examples of systems that either require or would benefit from a painted-potential approach.}
    \label{fig:paintedlattices}
\end{figure}

\section{Discussion}
\label{sec:discussion}

In this paper, we have derived expressions for the non-trivial Hubbard models
resulting from painted potentials. We allow different lattice sites to have different depths, widths and Rayleigh lengths, consistent with arbitrary basis or impurity problems. Key results are the hopping,
%
%\begin{widetext}
\begin{align*}
t_{{\rm deep},\ksite\lsite}=&\frac{2^{3/2}\sqrt{\wspota\wspotb}(\vzspota\vzspotb)^{3/8} \sqrt[4]{\zra \zrb} e^{-\frac{a^2 \sqrt{M \vzspota\vzspotb}}{\hbar W}}}{ W^3
   Z^{1/2}}\\
   & \hspace{20mm}\times\Bigg(2 a^2  \vzspota\vzspotb+(\vzspota+\vzspotb) W^2\Bigg)
\end{align*}
%\end{widetext}
%
and Hubbard $U$,
\begin{equation*}
  U^{\rm({\text{HO}})}_{\ksite\lsite,\rm Fesh} = g \frac{\left(M \vzspota\vzspotb\right)^{3/4} 2^{5/4}}{W Z^{1/2} \pi^{3/2} \hbar^{3/2}} e^{-2\frac{\sqrt{M\vzspota\vzspotb}  |\Rvecd_{\ksite\lsite}|^2}{W\hbar}}.
\end{equation*}
We note that these expressions would also be valid for bosons and Bose--Hubbard models. 

Table \ref{tab:qs} summarizes the components of the painted potential quantum simulator, and their correspondences with quantum simulators formed using sinusoidal optical lattices and condensed matter systems. A number of things are possible in painted potential systems that are not possible in purely sinusoidal lattices, particularly on a local (site) scale. Also, there are several additional ways to tune such a quantum simulator by changing spot depth, width, and spacing independently.

The ability to tune Hamiltonian parameters on the scale of individual sites means that painted potentials can be used for quantum simulation of non-trivial Hubbard models. Complicated
geometries can be implemented, with site-dependent Hubbard $U$ values. In
this way, painted potentials can be thought of as a toolkit for the
convenient implementation of custom models of strong correlation.

Problems that would benefit from the flexibility of the painted-potential approach fall broadly into two categories: systems where translational symmetry is broken, and systems with a complicated basis. Figure \ref{fig:paintedlattices} shows examples of condensed-matter problems that either require or would benefit from a painted-potential quantum simulator.

Defects and domain boundaries break translational symmetry and are therefore impossible to implement using only counterpropagating beams. The schematic in Figure \ref{fig:paintedlattices}(a) shows a Kondo-like defect in a one-dimensional chain (similar states can also be realised in two dimensions). For the quantum simulation of such systems, separate control over the defect site and the lattice is needed. Domain boundaries, e.g. highlighted in the dashed box on panel (b), are of interest in the study of the robustness of topological edge states. Again, loss of translational symmetry means such states are impossible to implement using only counter-propagating beams. From a condensed matter perspective, defects and impurities lead to qualitative changes in physics, such as the Kondo effect.

Lattices with complicated geometry are very challenging to implement with counter-propagating beams.  Topological states often occur in the vicinity of Dirac states, which may be found in bipartite lattices such as those in panels ($c$) and ($d$). Two-dimensional materials such as IV-VI semiconductors (similar to graphene) have topological insulator states. Such lattices are highly challenging to implement using counter-propagating beams \cite{li2016}. We expect better control using painted potentials. Cuprate superconductors are examples of low-dimensional systems with a unit cell containing a basis, in this case CuO$_{2}$ plaquettes. Studying such plaquettes requires a complicated 3-site basis in the unit cell, and independent control over Hubbard parameters of individual sites/bands which would be extremely challenging to control with counterpropagating beams. The multiple interacting bands that emerge are non-trivial, and an interesting direction for future study, particularly considering the high level of debate regarding the origins of cuprate superconductivity in the condensed matter community.

\section*{Acknowledgements}

This work was supported in part by CoSeC, the Computational Science Centre for Research Communities, through CCP9.

\bibstyle{unsrt}
\bibliography{paintedpotentialrefs}

\begin{thebibliography}{36}%
\makeatletter
\providecommand \@ifxundefined [1]{%
 \@ifx{#1\undefined}
}%
\providecommand \@ifnum [1]{%
 \ifnum #1\expandafter \@firstoftwo
 \else \expandafter \@secondoftwo
 \fi
}%
\providecommand \@ifx [1]{%
 \ifx #1\expandafter \@firstoftwo
 \else \expandafter \@secondoftwo
 \fi
}%
\providecommand \natexlab [1]{#1}%
\providecommand \enquote  [1]{``#1''}%
\providecommand \bibnamefont  [1]{#1}%
\providecommand \bibfnamefont [1]{#1}%
\providecommand \citenamefont [1]{#1}%
\providecommand \href@noop [0]{\@secondoftwo}%
\providecommand \href [0]{\begingroup \@sanitize@url \@href}%
\providecommand \@href[1]{\@@startlink{#1}\@@href}%
\providecommand \@@href[1]{\endgroup#1\@@endlink}%
\providecommand \@sanitize@url [0]{\catcode `\\12\catcode `\$12\catcode
  `\&12\catcode `\#12\catcode `\^12\catcode `\_12\catcode `\%12\relax}%
\providecommand \@@startlink[1]{}%
\providecommand \@@endlink[0]{}%
\providecommand \url  [0]{\begingroup\@sanitize@url \@url }%
\providecommand \@url [1]{\endgroup\@href {#1}{\urlprefix }}%
\providecommand \urlprefix  [0]{URL }%
\providecommand \Eprint [0]{\href }%
\providecommand \doibase [0]{https://doi.org/}%
\providecommand \selectlanguage [0]{\@gobble}%
\providecommand \bibinfo  [0]{\@secondoftwo}%
\providecommand \bibfield  [0]{\@secondoftwo}%
\providecommand \translation [1]{[#1]}%
\providecommand \BibitemOpen [0]{}%
\providecommand \bibitemStop [0]{}%
\providecommand \bibitemNoStop [0]{.\EOS\space}%
\providecommand \EOS [0]{\spacefactor3000\relax}%
\providecommand \BibitemShut  [1]{\csname bibitem#1\endcsname}%
\let\auto@bib@innerbib\@empty
%</preamble>
\bibitem [{\citenamefont {Henderson}\ \emph {et~al.}(2009)\citenamefont
  {Henderson}, \citenamefont {Ryu}, \citenamefont {MacCormick},\ and\
  \citenamefont {Boshier}}]{henderson2009}%
  \BibitemOpen
  \bibfield  {author} {\bibinfo {author} {\bibfnamefont {K.}~\bibnamefont
  {Henderson}}, \bibinfo {author} {\bibfnamefont {C.}~\bibnamefont {Ryu}},
  \bibinfo {author} {\bibfnamefont {C.}~\bibnamefont {MacCormick}},\ and\
  \bibinfo {author} {\bibfnamefont {M.}~\bibnamefont {Boshier}},\ }\bibfield
  {title} {\bibinfo {title} {{Experimental demonstration of painting arbitrary
  and dynamic potentials for Bose–Einstein condensates}},\ }\href@noop {}
  {\bibfield  {journal} {\bibinfo  {journal} {New J. Phys.}\ }\textbf {\bibinfo
  {volume} {11}},\ \bibinfo {pages} {043030} (\bibinfo {year}
  {2009})}\BibitemShut {NoStop}%
\bibitem [{\citenamefont {Ebadi}\ \emph {et~al.}(2021)\citenamefont {Ebadi},
  \citenamefont {Wang}, \citenamefont {Levine}, \citenamefont {Keesling},
  \citenamefont {Semeghini}, \citenamefont {Omran}, \citenamefont {Bluvstein},
  \citenamefont {Samajdar}, \citenamefont {Pichler}, \citenamefont {Ho},
  \citenamefont {Choi}, \citenamefont {Sachdev}, \citenamefont {Greiner},
  \citenamefont {Vuletic},\ and\ \citenamefont {Lukin}}]{ebadi2020}%
  \BibitemOpen
  \bibfield  {author} {\bibinfo {author} {\bibfnamefont {S.}~\bibnamefont
  {Ebadi}}, \bibinfo {author} {\bibfnamefont {T.}~\bibnamefont {Wang}},
  \bibinfo {author} {\bibfnamefont {H.}~\bibnamefont {Levine}}, \bibinfo
  {author} {\bibfnamefont {A.}~\bibnamefont {Keesling}}, \bibinfo {author}
  {\bibfnamefont {G.}~\bibnamefont {Semeghini}}, \bibinfo {author}
  {\bibfnamefont {A.}~\bibnamefont {Omran}}, \bibinfo {author} {\bibfnamefont
  {D.}~\bibnamefont {Bluvstein}}, \bibinfo {author} {\bibfnamefont
  {R.}~\bibnamefont {Samajdar}}, \bibinfo {author} {\bibfnamefont
  {H.}~\bibnamefont {Pichler}}, \bibinfo {author} {\bibfnamefont
  {W.}~\bibnamefont {Ho}}, \bibinfo {author} {\bibfnamefont {S.}~\bibnamefont
  {Choi}}, \bibinfo {author} {\bibfnamefont {S.}~\bibnamefont {Sachdev}},
  \bibinfo {author} {\bibfnamefont {M.}~\bibnamefont {Greiner}}, \bibinfo
  {author} {\bibfnamefont {V.}~\bibnamefont {Vuletic}},\ and\ \bibinfo {author}
  {\bibfnamefont {M.}~\bibnamefont {Lukin}},\ }\bibfield  {title} {\bibinfo
  {title} {Quantum phases of matter on a 256-atom programmable quantum
  simulator},\ }\href@noop {} {\bibfield  {journal} {\bibinfo  {journal}
  {Nature}\ }\textbf {\bibinfo {volume} {595}},\ \bibinfo {pages} {227}
  (\bibinfo {year} {2021})}\BibitemShut {NoStop}%
\bibitem [{\citenamefont {Nogrette}\ \emph {et~al.}(2014)\citenamefont
  {Nogrette}, \citenamefont {Labuhn}, \citenamefont {Ravets}, \citenamefont
  {Barredo}, \citenamefont {B\'eguin}, \citenamefont {Vernier}, \citenamefont
  {Lahaye},\ and\ \citenamefont {Browaeys}}]{nogrette2014}%
  \BibitemOpen
  \bibfield  {author} {\bibinfo {author} {\bibfnamefont {F.}~\bibnamefont
  {Nogrette}}, \bibinfo {author} {\bibfnamefont {H.}~\bibnamefont {Labuhn}},
  \bibinfo {author} {\bibfnamefont {S.}~\bibnamefont {Ravets}}, \bibinfo
  {author} {\bibfnamefont {D.}~\bibnamefont {Barredo}}, \bibinfo {author}
  {\bibfnamefont {L.}~\bibnamefont {B\'eguin}}, \bibinfo {author}
  {\bibfnamefont {A.}~\bibnamefont {Vernier}}, \bibinfo {author} {\bibfnamefont
  {T.}~\bibnamefont {Lahaye}},\ and\ \bibinfo {author} {\bibfnamefont
  {A.}~\bibnamefont {Browaeys}},\ }\bibfield  {title} {\bibinfo {title}
  {Single-atom trapping in holographic {2D} arrays of microtraps with arbitrary
  geometries},\ }\href {https://doi.org/10.1103/PhysRevX.4.021034} {\bibfield
  {journal} {\bibinfo  {journal} {Phys. Rev. X}\ }\textbf {\bibinfo {volume}
  {4}},\ \bibinfo {pages} {021034} (\bibinfo {year} {2014})}\BibitemShut
  {NoStop}%
\bibitem [{\citenamefont {Barredo}\ \emph {et~al.}(2016)\citenamefont
  {Barredo}, \citenamefont {de~Léséleuc}, \citenamefont {Lienhard},
  \citenamefont {Lahaye},\ and\ \citenamefont {Browaeys}}]{Barredo2016}%
  \BibitemOpen
  \bibfield  {author} {\bibinfo {author} {\bibfnamefont {D.}~\bibnamefont
  {Barredo}}, \bibinfo {author} {\bibfnamefont {S.}~\bibnamefont
  {de~Léséleuc}}, \bibinfo {author} {\bibfnamefont {V.}~\bibnamefont
  {Lienhard}}, \bibinfo {author} {\bibfnamefont {T.}~\bibnamefont {Lahaye}},\
  and\ \bibinfo {author} {\bibfnamefont {A.}~\bibnamefont {Browaeys}},\
  }\bibfield  {title} {\bibinfo {title} {An atom-by-atom assembler of
  defect-free arbitrary two-dimensional atomic arrays},\ }\href
  {https://doi.org/10.1126/science.aah3778} {\bibfield  {journal} {\bibinfo
  {journal} {Science}\ }\textbf {\bibinfo {volume} {354}},\ \bibinfo {pages}
  {1021} (\bibinfo {year} {2016})}\BibitemShut {NoStop}%
\bibitem [{\citenamefont {Barredo}\ \emph {et~al.}(2018)\citenamefont
  {Barredo}, \citenamefont {Lienhard}, \citenamefont {de~Léséleuc},
  \citenamefont {Lahaye},\ and\ \citenamefont {Browaeys}}]{Barredo2018}%
  \BibitemOpen
  \bibfield  {author} {\bibinfo {author} {\bibfnamefont {D.}~\bibnamefont
  {Barredo}}, \bibinfo {author} {\bibfnamefont {V.}~\bibnamefont {Lienhard}},
  \bibinfo {author} {\bibfnamefont {S.}~\bibnamefont {de~Léséleuc}}, \bibinfo
  {author} {\bibfnamefont {T.}~\bibnamefont {Lahaye}},\ and\ \bibinfo {author}
  {\bibfnamefont {A.}~\bibnamefont {Browaeys}},\ }\bibfield  {title} {\bibinfo
  {title} {Synthetic three-dimensional atomic structures assembled atom by
  atom},\ }\href {https://doi.org/10.1038/s41586-018-0450-2} {\bibfield
  {journal} {\bibinfo  {journal} {Nature}\ }\textbf {\bibinfo {volume} {561}},\
  \bibinfo {pages} {79 } (\bibinfo {year} {2018})}\BibitemShut {NoStop}%
\bibitem [{\citenamefont {Greiner}\ \emph {et~al.}(2002)\citenamefont
  {Greiner}, \citenamefont {Mandel}, \citenamefont {Esslinger}, \citenamefont
  {Hänsch},\ and\ \citenamefont {Bloch}}]{greiner2002}%
  \BibitemOpen
  \bibfield  {author} {\bibinfo {author} {\bibfnamefont {M.}~\bibnamefont
  {Greiner}}, \bibinfo {author} {\bibfnamefont {O.}~\bibnamefont {Mandel}},
  \bibinfo {author} {\bibfnamefont {T.}~\bibnamefont {Esslinger}}, \bibinfo
  {author} {\bibfnamefont {T.}~\bibnamefont {Hänsch}},\ and\ \bibinfo {author}
  {\bibfnamefont {I.}~\bibnamefont {Bloch}},\ }\bibfield  {title} {\bibinfo
  {title} {Quantum phase transition from a superfluid to a {Mott} insulator in
  a gas of ultracold atoms},\ }\href@noop {} {\bibfield  {journal} {\bibinfo
  {journal} {Nature}\ }\textbf {\bibinfo {volume} {415}},\ \bibinfo {pages}
  {39} (\bibinfo {year} {2002})}\BibitemShut {NoStop}%
\bibitem [{\citenamefont {Jördens}\ \emph {et~al.}(2008)\citenamefont
  {Jördens}, \citenamefont {Strohmaier}, \citenamefont {Günter},
  \citenamefont {Moritz},\ and\ \citenamefont {Esslinger}}]{jordens2008}%
  \BibitemOpen
  \bibfield  {author} {\bibinfo {author} {\bibfnamefont {R.}~\bibnamefont
  {Jördens}}, \bibinfo {author} {\bibfnamefont {N.}~\bibnamefont
  {Strohmaier}}, \bibinfo {author} {\bibfnamefont {K.}~\bibnamefont {Günter}},
  \bibinfo {author} {\bibfnamefont {H.}~\bibnamefont {Moritz}},\ and\ \bibinfo
  {author} {\bibfnamefont {T.}~\bibnamefont {Esslinger}},\ }\bibfield  {title}
  {\bibinfo {title} {A {Mott} insulator of fermionic atoms in an optical
  lattice},\ }\href@noop {} {\bibfield  {journal} {\bibinfo  {journal}
  {Nature}\ }\textbf {\bibinfo {volume} {455}},\ \bibinfo {pages} {204}
  (\bibinfo {year} {2008})}\BibitemShut {NoStop}%
\bibitem [{\citenamefont {Bloch}\ \emph {et~al.}(2008)\citenamefont {Bloch},
  \citenamefont {Dalibard},\ and\ \citenamefont {Zwerger}}]{bloch2008a}%
  \BibitemOpen
  \bibfield  {author} {\bibinfo {author} {\bibfnamefont {I.}~\bibnamefont
  {Bloch}}, \bibinfo {author} {\bibfnamefont {J.}~\bibnamefont {Dalibard}},\
  and\ \bibinfo {author} {\bibfnamefont {W.}~\bibnamefont {Zwerger}},\
  }\bibfield  {title} {\bibinfo {title} {Many-body physics with ultracold
  gases},\ }\href {https://doi.org/10.1103/RevModPhys.80.885} {\bibfield
  {journal} {\bibinfo  {journal} {Rev. Mod. Phys.}\ }\textbf {\bibinfo {volume}
  {80}},\ \bibinfo {pages} {885} (\bibinfo {year} {2008})}\BibitemShut
  {NoStop}%
\bibitem [{\citenamefont {Bednorz}\ and\ \citenamefont
  {Müller}(1986)}]{bednorz1986}%
  \BibitemOpen
  \bibfield  {author} {\bibinfo {author} {\bibfnamefont {J.}~\bibnamefont
  {Bednorz}}\ and\ \bibinfo {author} {\bibfnamefont {K.}~\bibnamefont
  {Müller}},\ }\bibfield  {title} {\bibinfo {title} {Possible high {$T_{c}$}
  superconductivity in the {Ba--La--Cu--O} system.},\ }\href@noop {} {\bibfield
   {journal} {\bibinfo  {journal} {Z. Physik B - Condensed Matter}\ }\textbf
  {\bibinfo {volume} {64}},\ \bibinfo {pages} {189} (\bibinfo {year}
  {1986})}\BibitemShut {NoStop}%
\bibitem [{\citenamefont {Novoselov}\ \emph {et~al.}(2004)\citenamefont
  {Novoselov}, \citenamefont {Geim}, \citenamefont {Morozov}, \citenamefont
  {Jiang}, \citenamefont {Zhang}, \citenamefont {Dubonos}, \citenamefont
  {Grigorieva},\ and\ \citenamefont {Firsov}}]{novoselov2004}%
  \BibitemOpen
  \bibfield  {author} {\bibinfo {author} {\bibfnamefont {K.}~\bibnamefont
  {Novoselov}}, \bibinfo {author} {\bibfnamefont {A.}~\bibnamefont {Geim}},
  \bibinfo {author} {\bibfnamefont {S.}~\bibnamefont {Morozov}}, \bibinfo
  {author} {\bibfnamefont {D.}~\bibnamefont {Jiang}}, \bibinfo {author}
  {\bibfnamefont {Y.}~\bibnamefont {Zhang}}, \bibinfo {author} {\bibfnamefont
  {S.}~\bibnamefont {Dubonos}}, \bibinfo {author} {\bibfnamefont
  {I.}~\bibnamefont {Grigorieva}},\ and\ \bibinfo {author} {\bibfnamefont
  {A.}~\bibnamefont {Firsov}},\ }\bibfield  {title} {\bibinfo {title} {Electric
  field effect in atomically thin carbon films},\ }\href@noop {} {\bibfield
  {journal} {\bibinfo  {journal} {Science}\ }\textbf {\bibinfo {volume}
  {306}},\ \bibinfo {pages} {666} (\bibinfo {year} {2004})}\BibitemShut
  {NoStop}%
\bibitem [{\citenamefont {Fallani}\ \emph {et~al.}(2007)\citenamefont
  {Fallani}, \citenamefont {Lye}, \citenamefont {Guarrera}, \citenamefont
  {Fort},\ and\ \citenamefont {Inguscio}}]{Fallani2007}%
  \BibitemOpen
  \bibfield  {author} {\bibinfo {author} {\bibfnamefont {L.}~\bibnamefont
  {Fallani}}, \bibinfo {author} {\bibfnamefont {J.~E.}\ \bibnamefont {Lye}},
  \bibinfo {author} {\bibfnamefont {V.}~\bibnamefont {Guarrera}}, \bibinfo
  {author} {\bibfnamefont {C.}~\bibnamefont {Fort}},\ and\ \bibinfo {author}
  {\bibfnamefont {M.}~\bibnamefont {Inguscio}},\ }\bibfield  {title} {\bibinfo
  {title} {Ultracold atoms in a disordered crystal of light: Towards a {Bose}
  glass},\ }\href
  {https://doi.org/https://doi.org/10.1103/PhysRevLett.98.130404} {\bibfield
  {journal} {\bibinfo  {journal} {Phys. Rev. Lett.}\ }\textbf {\bibinfo
  {volume} {98}},\ \bibinfo {pages} {130404} (\bibinfo {year}
  {2007})}\BibitemShut {NoStop}%
\bibitem [{\citenamefont {Roati}\ \emph {et~al.}(2008)\citenamefont {Roati},
  \citenamefont {D’Errico}, \citenamefont {L.} \emph {et~al.}}]{Roati2008}%
  \BibitemOpen
  \bibfield  {author} {\bibinfo {author} {\bibfnamefont {G.}~\bibnamefont
  {Roati}}, \bibinfo {author} {\bibfnamefont {C.}~\bibnamefont {D’Errico}},
  \bibinfo {author} {\bibfnamefont {F.}~\bibnamefont {L.}}, \emph {et~al.},\
  }\bibfield  {title} {\bibinfo {title} {Anderson localization of a
  non-interacting {Bose–Einstein} condensate.},\ }\href
  {https://doi.org/https://doi.org/10.1038/nature07071} {\bibfield  {journal}
  {\bibinfo  {journal} {Nature}\ }\textbf {\bibinfo {volume} {453}},\ \bibinfo
  {pages} {895–898} (\bibinfo {year} {2008})}\BibitemShut {NoStop}%
\bibitem [{\citenamefont {Sanchez-Palencia}\ and\ \citenamefont
  {Lewenstein}(2010)}]{SanchezPalencia2009}%
  \BibitemOpen
  \bibfield  {author} {\bibinfo {author} {\bibfnamefont {L.}~\bibnamefont
  {Sanchez-Palencia}}\ and\ \bibinfo {author} {\bibfnamefont {M.}~\bibnamefont
  {Lewenstein}},\ }\bibfield  {title} {\bibinfo {title} {Disordered quantum
  gases under control.},\ }\href
  {https://doi.org/https://doi.org/10.1038/nphys1507} {\bibfield  {journal}
  {\bibinfo  {journal} {Nature Phys.}\ }\textbf {\bibinfo {volume} {6}},\
  \bibinfo {pages} {87 } (\bibinfo {year} {2010})}\BibitemShut {NoStop}%
\bibitem [{\citenamefont {Andrei}\ \emph {et~al.}(1983)\citenamefont {Andrei},
  \citenamefont {Furuya},\ and\ \citenamefont {Lowenstein}}]{andrei1983}%
  \BibitemOpen
  \bibfield  {author} {\bibinfo {author} {\bibfnamefont {N.}~\bibnamefont
  {Andrei}}, \bibinfo {author} {\bibfnamefont {K.}~\bibnamefont {Furuya}},\
  and\ \bibinfo {author} {\bibfnamefont {J.~H.}\ \bibnamefont {Lowenstein}},\
  }\bibfield  {title} {\bibinfo {title} {Solution of the {Kondo} problem},\
  }\href {https://doi.org/10.1103/RevModPhys.55.331} {\bibfield  {journal}
  {\bibinfo  {journal} {Rev. Mod. Phys.}\ }\textbf {\bibinfo {volume} {55}},\
  \bibinfo {pages} {331} (\bibinfo {year} {1983})}\BibitemShut {NoStop}%
\bibitem [{\citenamefont {Eisert}\ \emph {et~al.}(2015)\citenamefont {Eisert},
  \citenamefont {Friesdorf},\ and\ \citenamefont {Gogolin}}]{eisert2015}%
  \BibitemOpen
  \bibfield  {author} {\bibinfo {author} {\bibfnamefont {J.}~\bibnamefont
  {Eisert}}, \bibinfo {author} {\bibfnamefont {M.}~\bibnamefont {Friesdorf}},\
  and\ \bibinfo {author} {\bibfnamefont {C.}~\bibnamefont {Gogolin}},\
  }\bibfield  {title} {\bibinfo {title} {Quantum many-body systems out of
  equilibrium},\ }\href@noop {} {\bibfield  {journal} {\bibinfo  {journal}
  {Nature Phys}\ }\textbf {\bibinfo {volume} {11}},\ \bibinfo {pages} {124}
  (\bibinfo {year} {2015})}\BibitemShut {NoStop}%
\bibitem [{\citenamefont {Hubbard}(1963)}]{hubbard1963}%
  \BibitemOpen
  \bibfield  {author} {\bibinfo {author} {\bibfnamefont {J.}~\bibnamefont
  {Hubbard}},\ }\bibfield  {title} {\bibinfo {title} {Electron correlations in
  narrow energy bands},\ }\href@noop {} {\bibfield  {journal} {\bibinfo
  {journal} {Proc. R. Soc. A}\ }\textbf {\bibinfo {volume} {276}},\ \bibinfo
  {pages} {238} (\bibinfo {year} {1963})}\BibitemShut {NoStop}%
\bibitem [{\citenamefont {Hirsch}(1984)}]{hirsch1984}%
  \BibitemOpen
  \bibfield  {author} {\bibinfo {author} {\bibfnamefont {J.~E.}\ \bibnamefont
  {Hirsch}},\ }\bibfield  {title} {\bibinfo {title} {Charge-density-wave to
  spin-density-wave transition in the extended {Hubbard} model},\ }\href
  {https://doi.org/10.1103/PhysRevLett.53.2327} {\bibfield  {journal} {\bibinfo
   {journal} {Phys. Rev. Lett.}\ }\textbf {\bibinfo {volume} {53}},\ \bibinfo
  {pages} {2327} (\bibinfo {year} {1984})}\BibitemShut {NoStop}%
\bibitem [{\citenamefont {Tarruell}\ \emph {et~al.}(2012)\citenamefont
  {Tarruell}, \citenamefont {Greif}, \citenamefont {Uehlinger} \emph
  {et~al.}}]{tarruell2012}%
  \BibitemOpen
  \bibfield  {author} {\bibinfo {author} {\bibfnamefont {L.}~\bibnamefont
  {Tarruell}}, \bibinfo {author} {\bibfnamefont {D.}~\bibnamefont {Greif}},
  \bibinfo {author} {\bibfnamefont {T.}~\bibnamefont {Uehlinger}}, \emph
  {et~al.},\ }\bibfield  {title} {\bibinfo {title} {Creating, moving and
  merging {Dirac} points with a {F}ermi gas in a tunable honeycomb lattice},\
  }\href@noop {} {\bibfield  {journal} {\bibinfo  {journal} {Nature}\ }\textbf
  {\bibinfo {volume} {483}},\ \bibinfo {pages} {302} (\bibinfo {year}
  {2012})}\BibitemShut {NoStop}%
\bibitem [{\citenamefont {Struck}\ \emph {et~al.}(2011)\citenamefont {Struck},
  \citenamefont {Ölschläger}, \citenamefont {Targat}, \citenamefont
  {Soltan-Panahi}, \citenamefont {Eckardt}, \citenamefont {Lewenstein},
  \citenamefont {Windpassinger},\ and\ \citenamefont {Sengstock}}]{struck2011}%
  \BibitemOpen
  \bibfield  {author} {\bibinfo {author} {\bibfnamefont {J.}~\bibnamefont
  {Struck}}, \bibinfo {author} {\bibfnamefont {C.}~\bibnamefont
  {Ölschläger}}, \bibinfo {author} {\bibfnamefont {R.~L.}\ \bibnamefont
  {Targat}}, \bibinfo {author} {\bibfnamefont {P.}~\bibnamefont
  {Soltan-Panahi}}, \bibinfo {author} {\bibfnamefont {A.}~\bibnamefont
  {Eckardt}}, \bibinfo {author} {\bibfnamefont {M.}~\bibnamefont {Lewenstein}},
  \bibinfo {author} {\bibfnamefont {P.}~\bibnamefont {Windpassinger}},\ and\
  \bibinfo {author} {\bibfnamefont {K.}~\bibnamefont {Sengstock}},\ }\bibfield
  {title} {\bibinfo {title} {Quantum simulation of frustrated classical
  magnetism in triangular optical lattices},\ }\href@noop {} {\bibfield
  {journal} {\bibinfo  {journal} {Science}\ }\textbf {\bibinfo {volume}
  {333}},\ \bibinfo {pages} {996} (\bibinfo {year} {2011})}\BibitemShut
  {NoStop}%
\bibitem [{\citenamefont {Jo}\ \emph {et~al.}(2012)\citenamefont {Jo},
  \citenamefont {Guzman}, \citenamefont {Thomas}, \citenamefont {Hosur},
  \citenamefont {Vishwanath},\ and\ \citenamefont {Stamper-Kurn}}]{jo2012}%
  \BibitemOpen
  \bibfield  {author} {\bibinfo {author} {\bibfnamefont {G.-B.}\ \bibnamefont
  {Jo}}, \bibinfo {author} {\bibfnamefont {J.}~\bibnamefont {Guzman}}, \bibinfo
  {author} {\bibfnamefont {C.~K.}\ \bibnamefont {Thomas}}, \bibinfo {author}
  {\bibfnamefont {P.}~\bibnamefont {Hosur}}, \bibinfo {author} {\bibfnamefont
  {A.}~\bibnamefont {Vishwanath}},\ and\ \bibinfo {author} {\bibfnamefont
  {D.~M.}\ \bibnamefont {Stamper-Kurn}},\ }\bibfield  {title} {\bibinfo {title}
  {Ultracold atoms in a tunable optical {Kagome} lattice},\ }\href
  {https://doi.org/10.1103/PhysRevLett.108.045305} {\bibfield  {journal}
  {\bibinfo  {journal} {Phys. Rev. Lett.}\ }\textbf {\bibinfo {volume} {108}},\
  \bibinfo {pages} {045305} (\bibinfo {year} {2012})}\BibitemShut {NoStop}%
\bibitem [{\citenamefont {Sebby-Strabley}\ \emph {et~al.}(2006)\citenamefont
  {Sebby-Strabley}, \citenamefont {Anderlini}, \citenamefont {Jessen},\ and\
  \citenamefont {Porto}}]{sebbystrabley2006}%
  \BibitemOpen
  \bibfield  {author} {\bibinfo {author} {\bibfnamefont {J.}~\bibnamefont
  {Sebby-Strabley}}, \bibinfo {author} {\bibfnamefont {M.}~\bibnamefont
  {Anderlini}}, \bibinfo {author} {\bibfnamefont {P.~S.}\ \bibnamefont
  {Jessen}},\ and\ \bibinfo {author} {\bibfnamefont {J.~V.}\ \bibnamefont
  {Porto}},\ }\bibfield  {title} {\bibinfo {title} {Lattice of double wells for
  manipulating pairs of cold atoms},\ }\href
  {https://doi.org/10.1103/PhysRevA.73.033605} {\bibfield  {journal} {\bibinfo
  {journal} {Phys. Rev. A}\ }\textbf {\bibinfo {volume} {73}},\ \bibinfo
  {pages} {033605} (\bibinfo {year} {2006})}\BibitemShut {NoStop}%
\bibitem [{\citenamefont {Goldman}\ \emph {et~al.}(2016)\citenamefont
  {Goldman}, \citenamefont {Budich},\ and\ \citenamefont
  {Zoller}}]{goldman2016}%
  \BibitemOpen
  \bibfield  {author} {\bibinfo {author} {\bibfnamefont {N.}~\bibnamefont
  {Goldman}}, \bibinfo {author} {\bibfnamefont {J.}~\bibnamefont {Budich}},\
  and\ \bibinfo {author} {\bibfnamefont {P.}~\bibnamefont {Zoller}},\
  }\bibfield  {title} {\bibinfo {title} {Topological quantum matter with
  ultracold gases in optical lattices},\ }\href@noop {} {\bibfield  {journal}
  {\bibinfo  {journal} {Nature Phys.}\ }\textbf {\bibinfo {volume} {12}},\
  \bibinfo {pages} {639} (\bibinfo {year} {2016})}\BibitemShut {NoStop}%
\bibitem [{\citenamefont {Kondo}(1964)}]{kondo1964}%
  \BibitemOpen
  \bibfield  {author} {\bibinfo {author} {\bibfnamefont {J.}~\bibnamefont
  {Kondo}},\ }\bibfield  {title} {\bibinfo {title} {Resistance minimum in
  dilute magnetic alloys},\ }\href@noop {} {\bibfield  {journal} {\bibinfo
  {journal} {Progress of Theoretical Physics}\ }\textbf {\bibinfo {volume}
  {32}},\ \bibinfo {pages} {37} (\bibinfo {year} {1964})}\BibitemShut {NoStop}%
\bibitem [{\citenamefont {Jaksch}\ \emph {et~al.}(1998)\citenamefont {Jaksch},
  \citenamefont {Bruder}, \citenamefont {Cirac}, \citenamefont {Gardiner},\
  and\ \citenamefont {Zoller}}]{jaksch1998}%
  \BibitemOpen
  \bibfield  {author} {\bibinfo {author} {\bibfnamefont {D.}~\bibnamefont
  {Jaksch}}, \bibinfo {author} {\bibfnamefont {C.}~\bibnamefont {Bruder}},
  \bibinfo {author} {\bibfnamefont {J.~I.}\ \bibnamefont {Cirac}}, \bibinfo
  {author} {\bibfnamefont {C.~W.}\ \bibnamefont {Gardiner}},\ and\ \bibinfo
  {author} {\bibfnamefont {P.}~\bibnamefont {Zoller}},\ }\bibfield  {title}
  {\bibinfo {title} {Cold bosonic atoms in optical lattices},\ }\href
  {https://doi.org/10.1103/PhysRevLett.81.3108} {\bibfield  {journal} {\bibinfo
   {journal} {Phys. Rev. Lett.}\ }\textbf {\bibinfo {volume} {81}},\ \bibinfo
  {pages} {3108} (\bibinfo {year} {1998})}\BibitemShut {NoStop}%
\bibitem [{\citenamefont {Wall}\ \emph {et~al.}(2015)\citenamefont {Wall},
  \citenamefont {Hazzard},\ and\ \citenamefont {Rey}}]{wall2015}%
  \BibitemOpen
  \bibfield  {author} {\bibinfo {author} {\bibfnamefont {M.~L.}\ \bibnamefont
  {Wall}}, \bibinfo {author} {\bibfnamefont {K.~R.~A.}\ \bibnamefont
  {Hazzard}},\ and\ \bibinfo {author} {\bibfnamefont {A.~M.}\ \bibnamefont
  {Rey}},\ }\bibfield  {title} {\bibinfo {title} {Effective many-body
  parameters for atoms in nonseparable gaussian optical potentials},\ }\href
  {https://doi.org/10.1103/PhysRevA.92.013610} {\bibfield  {journal} {\bibinfo
  {journal} {Phys. Rev. A}\ }\textbf {\bibinfo {volume} {92}},\ \bibinfo
  {pages} {013610} (\bibinfo {year} {2015})}\BibitemShut {NoStop}%
\bibitem [{\citenamefont {Grimm}\ \emph {et~al.}(2000)\citenamefont {Grimm},
  \citenamefont {Weidemüller},\ and\ \citenamefont {Ovchinnikov}}]{grimm2000}%
  \BibitemOpen
  \bibfield  {author} {\bibinfo {author} {\bibfnamefont {R.}~\bibnamefont
  {Grimm}}, \bibinfo {author} {\bibfnamefont {M.}~\bibnamefont
  {Weidemüller}},\ and\ \bibinfo {author} {\bibfnamefont {Y.~B.}\ \bibnamefont
  {Ovchinnikov}},\ }\bibfield  {title} {\bibinfo {title} {Optical dipole traps
  for neutral atoms},\ }\href@noop {} {\bibfield  {journal} {\bibinfo
  {journal} {Advances in Atomic, Molecular and Optical Physics}\ }\textbf
  {\bibinfo {volume} {42}},\ \bibinfo {pages} {95} (\bibinfo {year}
  {2000})}\BibitemShut {NoStop}%
\bibitem [{\citenamefont {Hague}\ and\ \citenamefont
  {MacCormick}(2012)}]{hague2012a}%
  \BibitemOpen
  \bibfield  {author} {\bibinfo {author} {\bibfnamefont {J.~P.}\ \bibnamefont
  {Hague}}\ and\ \bibinfo {author} {\bibfnamefont {C.}~\bibnamefont
  {MacCormick}},\ }\bibfield  {title} {\bibinfo {title} {Bilayers of rydberg
  atoms as a quantum simulator for unconventional superconductors},\
  }\href@noop {} {\bibfield  {journal} {\bibinfo  {journal} {Phy. Rev. Lett.}\
  }\textbf {\bibinfo {volume} {109}},\ \bibinfo {pages} {223001} (\bibinfo
  {year} {2012})}\BibitemShut {NoStop}%
\bibitem [{\citenamefont {Ray}(1992)}]{ray}%
  \BibitemOpen
  \bibfield  {author} {\bibinfo {author} {\bibfnamefont {A.}~\bibnamefont
  {Ray}},\ }\href@noop {} {\emph {\bibinfo {title} {Quantum mechanics}}},\
  \bibinfo {edition} {3rd}\ ed.\ (\bibinfo  {publisher} {IOP publishing},\
  \bibinfo {year} {1992})\BibitemShut {NoStop}%
\bibitem [{Note1()}]{Note1}%
  \BibitemOpen
  \bibinfo {note} {This argument would not work with Coulomb lattice
  potentials, as they have long range tails, however there are no long range
  tails on spot potentials allowing truncation of the sum}\BibitemShut
  {NoStop}%
\bibitem [{Note2()}]{Note2}%
  \BibitemOpen
  \bibinfo {note} {We note that this argument would not work for an unscreened
  Coulomb potential in a traditional condensed matter tight-binding
  approximation.}\BibitemShut {Stop}%
\bibitem [{\citenamefont {Gill}(1994)}]{gill1994a}%
  \BibitemOpen
  \bibfield  {author} {\bibinfo {author} {\bibfnamefont {P.~M.~W.}\
  \bibnamefont {Gill}},\ }\bibfield  {title} {\bibinfo {title} {Molecular
  integrals over {G}aussian basis functions},\ }\href@noop {} {\bibfield
  {journal} {\bibinfo  {journal} {Advances in quantum chemistry}\ }\textbf
  {\bibinfo {volume} {25}},\ \bibinfo {pages} {141} (\bibinfo {year}
  {1994})}\BibitemShut {NoStop}%
\bibitem [{\citenamefont {Schunck}\ \emph {et~al.}(2005)\citenamefont
  {Schunck}, \citenamefont {Zwierlein}, \citenamefont {Stan}, \citenamefont
  {Raupach}, \citenamefont {Ketterle}, \citenamefont {Simoni}, \citenamefont
  {Tiesinga}, \citenamefont {Williams},\ and\ \citenamefont
  {Julienne}}]{Schunck2005}%
  \BibitemOpen
  \bibfield  {author} {\bibinfo {author} {\bibfnamefont {C.~H.}\ \bibnamefont
  {Schunck}}, \bibinfo {author} {\bibfnamefont {M.~W.}\ \bibnamefont
  {Zwierlein}}, \bibinfo {author} {\bibfnamefont {C.~A.}\ \bibnamefont {Stan}},
  \bibinfo {author} {\bibfnamefont {S.~M.~F.}\ \bibnamefont {Raupach}},
  \bibinfo {author} {\bibfnamefont {W.}~\bibnamefont {Ketterle}}, \bibinfo
  {author} {\bibfnamefont {A.}~\bibnamefont {Simoni}}, \bibinfo {author}
  {\bibfnamefont {E.}~\bibnamefont {Tiesinga}}, \bibinfo {author}
  {\bibfnamefont {C.~J.}\ \bibnamefont {Williams}},\ and\ \bibinfo {author}
  {\bibfnamefont {P.~S.}\ \bibnamefont {Julienne}},\ }\bibfield  {title}
  {\bibinfo {title} {Feshbach resonances in fermionic $^{6}\mathrm{Li}$},\
  }\href {https://doi.org/10.1103/PhysRevA.71.045601} {\bibfield  {journal}
  {\bibinfo  {journal} {Phys. Rev. A}\ }\textbf {\bibinfo {volume} {71}},\
  \bibinfo {pages} {045601} (\bibinfo {year} {2005})}\BibitemShut {NoStop}%
\bibitem [{\citenamefont {Sortais}\ \emph {et~al.}(2007)\citenamefont
  {Sortais}, \citenamefont {Marion}, \citenamefont {Tuchendler}, \citenamefont
  {Lance}, \citenamefont {Lamare}, \citenamefont {Fournet}, \citenamefont
  {Armellin}, \citenamefont {Mercier}, \citenamefont {Messin}, \citenamefont
  {Browaeys},\ and\ \citenamefont {Grangier}}]{sortais2007}%
  \BibitemOpen
  \bibfield  {author} {\bibinfo {author} {\bibfnamefont {Y.~R.~P.}\
  \bibnamefont {Sortais}}, \bibinfo {author} {\bibfnamefont {H.}~\bibnamefont
  {Marion}}, \bibinfo {author} {\bibfnamefont {C.}~\bibnamefont {Tuchendler}},
  \bibinfo {author} {\bibfnamefont {A.~M.}\ \bibnamefont {Lance}}, \bibinfo
  {author} {\bibfnamefont {M.}~\bibnamefont {Lamare}}, \bibinfo {author}
  {\bibfnamefont {P.}~\bibnamefont {Fournet}}, \bibinfo {author} {\bibfnamefont
  {C.}~\bibnamefont {Armellin}}, \bibinfo {author} {\bibfnamefont
  {R.}~\bibnamefont {Mercier}}, \bibinfo {author} {\bibfnamefont
  {G.}~\bibnamefont {Messin}}, \bibinfo {author} {\bibfnamefont
  {A.}~\bibnamefont {Browaeys}},\ and\ \bibinfo {author} {\bibfnamefont
  {P.}~\bibnamefont {Grangier}},\ }\bibfield  {title} {\bibinfo {title}
  {Diffraction-limited optics for single-atom manipulation},\ }\href@noop {}
  {\bibfield  {journal} {\bibinfo  {journal} {Phys. Rev. A}\ }\textbf {\bibinfo
  {volume} {75}},\ \bibinfo {pages} {013406} (\bibinfo {year}
  {2007})}\BibitemShut {NoStop}%
\bibitem [{\citenamefont {Bakr}\ \emph {et~al.}(2009)\citenamefont {Bakr},
  \citenamefont {Gillen}, \citenamefont {Peng}, \citenamefont {F\"olling},\
  and\ \citenamefont {Greiner}}]{Bakr2009}%
  \BibitemOpen
  \bibfield  {author} {\bibinfo {author} {\bibfnamefont {W.}~\bibnamefont
  {Bakr}}, \bibinfo {author} {\bibfnamefont {J.}~\bibnamefont {Gillen}},
  \bibinfo {author} {\bibfnamefont {A.}~\bibnamefont {Peng}}, \bibinfo {author}
  {\bibfnamefont {S.}~\bibnamefont {F\"olling}},\ and\ \bibinfo {author}
  {\bibfnamefont {M.}~\bibnamefont {Greiner}},\ }\bibfield  {title} {\bibinfo
  {title} {A quantum gas microscope for detecting single atoms in a
  {Hubbard}-regime optical lattice},\ }\href
  {https://doi.org/https://doi.org/10.1038/nature08482} {\bibfield  {journal}
  {\bibinfo  {journal} {Nature}\ }\textbf {\bibinfo {volume} {462}},\ \bibinfo
  {pages} {74} (\bibinfo {year} {2009})}\BibitemShut {NoStop}%
\bibitem [{\citenamefont {Keller}\ \emph {et~al.}(2014)\citenamefont {Keller},
  \citenamefont {Kotyrba}, \citenamefont {Leupold}, \citenamefont {Singh},
  \citenamefont {Ebner},\ and\ \citenamefont {Zeilinger}}]{Keller2014}%
  \BibitemOpen
  \bibfield  {author} {\bibinfo {author} {\bibfnamefont {M.}~\bibnamefont
  {Keller}}, \bibinfo {author} {\bibfnamefont {M.}~\bibnamefont {Kotyrba}},
  \bibinfo {author} {\bibfnamefont {F.}~\bibnamefont {Leupold}}, \bibinfo
  {author} {\bibfnamefont {M.}~\bibnamefont {Singh}}, \bibinfo {author}
  {\bibfnamefont {M.}~\bibnamefont {Ebner}},\ and\ \bibinfo {author}
  {\bibfnamefont {A.}~\bibnamefont {Zeilinger}},\ }\bibfield  {title} {\bibinfo
  {title} {{Bose-Einstein} condensate of metastable helium for quantum
  correlation experiments},\ }\href@noop {} {\bibfield  {journal} {\bibinfo
  {journal} {Phys. Rev. A}\ }\textbf {\bibinfo {volume} {90}},\ \bibinfo
  {pages} {063607} (\bibinfo {year} {2014})}\BibitemShut {NoStop}%
\bibitem [{\citenamefont {Li}\ \emph {et~al.}(2016)\citenamefont {Li},
  \citenamefont {Duca}, \citenamefont {Reitter}, \citenamefont {Grusdt},
  \citenamefont {Demler}, \citenamefont {Endres}, \citenamefont
  {Schleier-Smith}, \citenamefont {Bloch},\ and\ \citenamefont
  {Schneider}}]{li2016}%
  \BibitemOpen
  \bibfield  {author} {\bibinfo {author} {\bibfnamefont {T.}~\bibnamefont
  {Li}}, \bibinfo {author} {\bibfnamefont {L.}~\bibnamefont {Duca}}, \bibinfo
  {author} {\bibfnamefont {M.}~\bibnamefont {Reitter}}, \bibinfo {author}
  {\bibfnamefont {F.}~\bibnamefont {Grusdt}}, \bibinfo {author} {\bibfnamefont
  {E.}~\bibnamefont {Demler}}, \bibinfo {author} {\bibfnamefont
  {M.}~\bibnamefont {Endres}}, \bibinfo {author} {\bibfnamefont
  {M.}~\bibnamefont {Schleier-Smith}}, \bibinfo {author} {\bibfnamefont
  {I.}~\bibnamefont {Bloch}},\ and\ \bibinfo {author} {\bibfnamefont
  {U.}~\bibnamefont {Schneider}},\ }\bibfield  {title} {\bibinfo {title} {Bloch
  state tomography using wilson lines},\ }\href@noop {} {\bibfield  {journal}
  {\bibinfo  {journal} {Science}\ }\textbf {\bibinfo {volume} {352}},\ \bibinfo
  {pages} {1094} (\bibinfo {year} {2016})}\BibitemShut {NoStop}%
\end{thebibliography}%

\appendix*

\section{Notation}
%\label{sec:appendix}

In Table \ref{tab:notation}, we provide a summary of all symbols used in this paper, to assist with the extensive notation. 

\begin{table*}[h]
    \caption{Summary of notation.}
    \centering
        \begin{tabular}{|l|l|l|l|}
    \hline
    \multicolumn{2}{|c|}{\textbf{Optical lattice parameters}} & \multicolumn{2}{|c|}{\textbf{Atom and laser parameters}}\\
    \hline
        $a$ & intersite distance & $B$ & applied magnetic field\\
        $\ksite,\lsite$ & site indices & $\Delta B$ & width of Feshbach resonance\\
        $\vzspot$ & magnitude of spot potential & $\Gamma$ &  transition lifetime\\
        $\wspot$ & spot waist & $\omega_{0}$ & transition frequency \\
        $\wspota,\wspotb$ & spot waist on sites $\ksite,\lsite$  & $\lambda_{0}$ & transition wavelength \\        

        $\bar{{\rm w}}(z)$ & spatial dependence of waist & $\lambda_\mathrm{D1}, \lambda_\mathrm{D2}$ &  D1 and D2 transition wavelengths \\
                $\zr$ & Rayleigh length & $\Delta$ & detuning from transition \\
                        $\zra,\zrb$ & Rayleigh length on sites $\ksite,\lsite$ & $I(\rvec)$ & laser intensity \\
        $\wpan$ & pancake waist & $I_{\rm sat}$ & saturation intensity \\
        $V_{\rm pan}(z)$ & pancake potential & $P$ & laser power\\
        $V_{\rm spot,i}(\rvecd,z)$ & spot potential & $M$ & atom mass\\
        $\tilde{V}(\rvecd,z)$ & z-axis Taylor expanded lattice potential & $g$ & interaction coupling constant\\
        $V_{site}(\rvecd,z)$ & all-axis Taylor expanded lattice potential & $\lambda_{\rm Las}$ & laser wavelength\\
$\vzpan$ & magnitude of pancake potential & $a_{\rm bg}$ & s-wave scattering length\\
        $\vzspota,\vzspotb$ & magnitude of spot potential on sites $\ksite,\lsite$ & $a_{s}$ & scattering length near Feshbach resonance\\
        $E_{R}$ & recoil energy & & \\
    \hline
    \multicolumn{2}{|c|}{\textbf{Hubbard Hamiltonian parameters}} & \multicolumn{2}{|c|}{\textbf{Coordinates}}\\
    \hline
        $t_{\lsite\ksite}$ & Hopping & $\rvecd$ & vector in plane of pancake\\
      $U$   & Hubbard $U$ & $\Rvecd$ & spot positions in plane of pancake\\
      $U_{\rm Fesh}$ & Hubbard $U$ derived from Feshbach resonance & $x$,$y$ & spatial coordinates in plane of pancake\\ 
        $E_{\lsite,{\text{HO}}}$ & energy of isolated site $\lsite$ & $z$ & spatial coordinate perpendicular to pancake\\
        $T_{\lsite\ksite}$ & overlap integral for spot potential & &\\
        $t_{\text{deep}}$ & hopping in deep well limit & & \\
        $U^{(HO)}$ & Hubbard $U$ in the deep well limit & & \\
    \hline
    \end{tabular}
    \label{tab:notation}
\end{table*}

\end{document}